\documentclass[%
 reprint,
 amsmath,amssymb,
 aps,
pra,
]{revtex4-2}

\usepackage{braket}
\usepackage{amsmath}
\usepackage{subfigure}
\usepackage{enumitem}
\usepackage{hyperref}%
\usepackage[normalem]{ulem}  %
\usepackage{adjustbox}
\usepackage{multirow}
\usepackage{booktabs}
\usepackage{comment}
\usepackage{graphicx}
\usepackage{algpseudocode}
\usepackage{algorithm}

\newcommand{\mat}[1]{\mathbf{#1}}
\newcommand{\vect}[1]{\boldsymbol{#1}}
\def\R{\mathbb{R}}
\def\E{\mathbb{E}}

\newcommand{\Gate}[1]{\textsc{#1}}

\newcommand{\zgate}{\Gate{Z}}

\makeatletter
\newcommand{\vast}{\bBigg@{3}}
\newcommand{\Vast}{\bBigg@{4}}
\makeatother

\DeclareMathOperator{\Tr}{Tr}
\def\R{\mathbb{R}}

\newcommand{\argmin}{\mathop{\mathrm{argmin}}}
\newcommand{\argmax}{\mathop{\mathrm{argmax}}}

\newcommand{\Rho}{\boldsymbol{\mathrm{P}}}

\newcommand{\rev}[1]{\textcolor{black}{#1}}

\usepackage{xcolor}
\begin{document}

\title{Distributionally Robust Variational Quantum Algorithms with Shifted Noise} 

\author{Zichang He\textsuperscript{1}, Bo Peng\textsuperscript{2}, Yuri Alexeev\textsuperscript{3} and Zheng Zhang\textsuperscript{1}}
\affiliation{
 \textsuperscript{1}Department of Electrical and Computer Engineering, University of California, Santa Barbara, Santa Barbara, CA 93106 USA
}
\affiliation{\textsuperscript{2}Physical and Computational Sciences Directorate, Pacific Northwest National Laboratory, Richland, WA, 99352 USA}
\affiliation{\textsuperscript{3}Computational Science Division, Argonne National Laboratory, Lemont, IL, 60439 USA}

\begin{abstract}
Given their potential to demonstrate near-term quantum advantage, variational quantum algorithms (VQAs) have been extensively studied. Although numerous techniques have been developed for VQA parameter optimization, it remains a significant challenge. \rev{A practical issue is that quantum noise is highly unstable and thus it is likely to shift in real time.} This presents a critical problem as an optimized VQA ansatz may not perform effectively under a different noise environment.
For the first time, we explore how to optimize VQA parameters to be robust against unknown shifted noise. We model the noise level as a random variable with an unknown probability density function (PDF), and we assume that the PDF may shift within an uncertainty set. This assumption guides us to formulate a distributionally robust optimization problem, with the goal of finding parameters that maintain effectiveness under shifted noise.
We utilize a distributionally robust Bayesian optimization solver for our proposed formulation. This provides numerical evidence in both the Quantum Approximate Optimization Algorithm (QAOA) and the Variational Quantum Eigensolver (VQE) with hardware-efficient ansatz, indicating that we can identify parameters that perform more robustly under shifted noise.
We regard this work as the first step towards improving the reliability of VQAs influenced by \rev{shifted noise from the parameter optimization perspective}.
\end{abstract}
\maketitle

\section{Introduction}
Variational quantum algorithms (VQAs)~\cite{cerezo2021variational} have the potential to demonstrate quantum advantage and have been applied in diverse fields, such as optimization~\cite{Moll_2018,egger2021warm}, finance~\cite{he2023alignment,orus2019quantum,herman2022survey}, machine learning~\cite{biamonte2017quantum,cerezo2022challenges,liu2023towards}, quantum simulation~\cite{miessen2023quantum,peng2022quantum,gulania2022quybe}, and chemistry~\cite{fedorov2022vqe, mcardle2020quantum,cao2019quantum}. However, parameter optimization is a substantial challenge for VQAs~\cite{bittel2021training}.

Numerous efforts have been made to optimize VQA parameters~\cite{sung2020using, bonet2023performance,gilyen2019optimizing,shaydulin2023parameter}. One critical challenge for VQA parameter optimization is quantum noise~\cite{stilck2021limitations,gonzalez2022error,de2023limitations}, which limits their capabilities and introduces additional complexities to parameter optimization. Modeling and mitigating hardware noise is a core part of Near-term Intermediate-scale Quantum (NISQ) algorithms~\cite{harper2020efficient, endo2018practical, suzuki2022quantum}. Quantifying and improving the reliability and robustness of a VQA has been an important task and has gained increasing attention recently. To name a few, machine learning methods have been used to estimate the reliability of a quantum circuit~\cite{liu2020reliability}; noise-aware ansatz design methodologies~\cite{wang2022quantumnas} and robust circuit realization from a lower-level abstraction~\cite{magann2021pulses,liang2022hybrid} have also been investigated. 

A more challenging yet practical problem is the \rev{instability of quantum noise}. Suppose we have an accurate model of the quantum noise as a reference. However, the quantum noise can change significantly under different environmental conditions in real time, making the reference noise model inaccurate. \rev{\cite{burnett2019decoherence} has shown that the noise fluctuation is usually less disinclined.}
Some studies~\cite{dasgupta2022assessing,dasgupta2022characterizing} have considered the reproducibility and stability under different noise models. We refer to this phenomenon of noise change as ``noise shift''. 
The optimization of variational quantum circuits and error mitigation under real-time noise has gained attention recently~\cite{hu2023toward,zhang2023disq,dasgupta2023reliable,dasgupta2023adaptive,ravi2023navigating}. 

In this paper, we ask a fundamental question: \emph{Can we optimize the VQA parameters such that they are robust to potentially shifted (unknown) noise?} 
We assume that we have access to a fixed noise model, but the actual noise level is an unknown random variable with an unknown PDF. This fixed PDF represents our limited knowledge about the potential noise shift. 

\begin{figure*}[t]
    \centering
    \includegraphics[width = 0.85\linewidth]{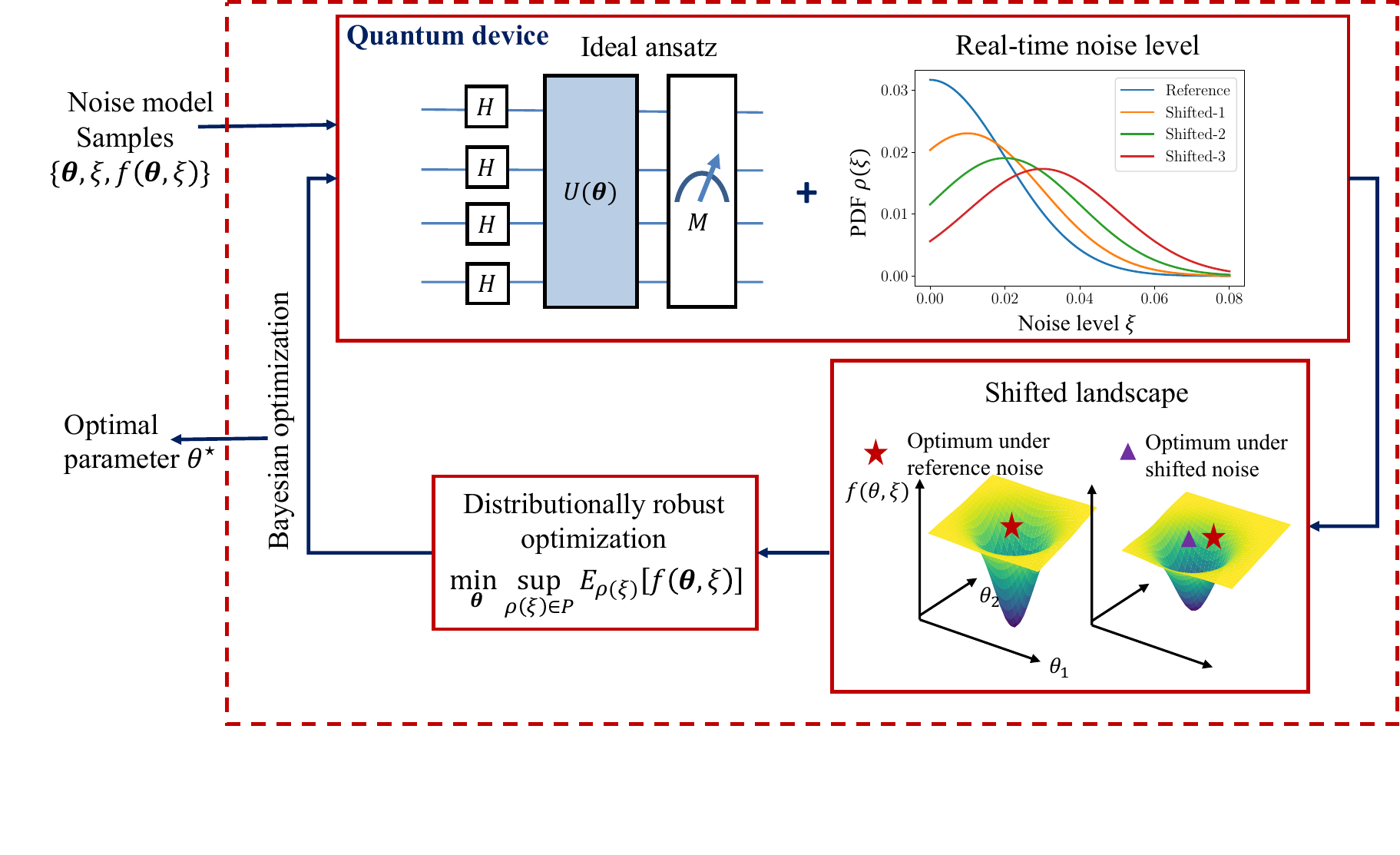}
    \caption{
    Overview of the distributionally robust variational quantum algorithms. Given an ideal ansatz and noise model, we assume the noise level is a random variable that can change in real time. We have samples of the noise level variable $\xi$ from a reference distribution, ansatz parameter $\vect{\theta}$, and the corresponding VQA performance $f(\vect{\theta},\xi)$. With the \rev{shifted} noise, the VQA landscape and its optimum $\vect{\theta}$ can potentially change. Specifically, the optimal $\vect{\theta}$ under a certain noise level may not perform well under another noise level. Likewise, an optimal $\vect{\theta}$ under a reference noise level PDF may not perform well under another noise level PDF. To address the landscape shift, we reformulate the parameter optimization problem as a min-max formulation to find a robust parameter $\vect{\theta}$. In other words, we aim to optimize the performance under the worst-case noise level PDF. We use a distributionally robust Bayesian optimization solver to solve the new parameter optimization formulation, which is still handled by classical computers.}
    \label{fig:intro_figure}
\end{figure*}

To optimize VQA parameters under such unknown noise, for the first time, we propose a new min-max optimization formulation. Such an optimization formulation is called distributionally robust optimization (DRO) in the classical operation research community~\cite{rahimian2019distributionally,lin2022distributionally,kuhn2019wasserstein,delage2010distributionally}. 
DRO is an advanced optimization framework that aims to find solutions resilient against a range of possible probability distributions rather than a single expected distribution.
In our context, we aim to optimize parameters \rev{against} the worst-case distribution of noise levels. This task, while distinct, complements error mitigation efforts. Rather than attempting to reduce quantum noise,  our approach focuses on optimizing parameters in the presence of potentially shifting noise. \rev{Furthermore, our method allows for seamless integration with various error mitigation techniques.}

\textbf{Paper contributions.}
In this work, we investigate the problem formulation, numerical solver, and validation of variational quantum algorithm training under unknown shifted noise. %
The overview is illustrated in Fig.~\ref{fig:intro_figure}. Our specific contributions include:
\begin{itemize}%
    \item 
    To be robust against the \rev{shifted} noise, we formulate the problem of optimizing VQA algorithms as a distributionally robust optimization that aims to optimize a targeted performance under the worst-case noise distribution.
    \rev{We characterize the quantum noise using a fixed noise model with uncertain and varying levels of strengths, where the noise level is a random variable with an unknown probability density function (PDF).}
    \item To solve the distributionally robust optimization, we model the unknown PDF as a distributional uncertainty set that is defined by Maximum Mean Discrepancy (MMD). We then solve the min-max problem utilizing a distributionally robust Bayesian optimization (DRBO) method~\cite{kirschner2020distributionally,tay2022efficient,husain2022distributionally}.
    Recently, Bayesian optimization (BO) has attracted attention in the field of quantum algorithm optimization~\cite{iannelli2021noisy,duffield2023bayesian,self2021variational,muller2022accelerating,tibaldi2023bayesian,finvzgar2024designing,tamiya2022stochastic, kim2023quantum, benitez2024bayesian, ravi2022cafqa, cheng2024quantum}. 
    \item We validate the proposed min-max formulation on two \rev{well-recognized} VQAs, \rev{namely} QAOA for the MaxCut problem and VQE with hardware-efficient ansatz for the one-dimensional Heisenberg model. \rev{Numerical} results show that the proposed parameter optimization algorithm \rev{outperforms conventional methods under shifted noise conditions}.
\end{itemize}

\section{Problem Formulation} 
Variational quantum algorithms (VQAs) are a class of algorithms in quantum computing that utilize a hybrid approach, combining classical and quantum computing resources to solve computational problems. They are especially pertinent for use with Noisy Intermediate-Scale Quantum (NISQ) devices, which are the currently available quantum hardware.

The core idea of VQAs is to define a parameterized quantum circuit (ansatz) that manipulates the state of a quantum system in a way that depends on a set of parameters $\boldsymbol{\theta}$. These parameters are then optimized classically to minimize an objective function $\braket{\vect{\psi}(\boldsymbol{\theta}) \vert \mat{O} \vert \vect{\psi}(\boldsymbol{\theta})}$, \rev{where $\vect{\psi}(\boldsymbol{\theta})$ is the resulting noiseless quantum state from the parameterized anstaz, $\mat{O}$ is an observable of interest and the objective function} is evaluated by the quantum system. However, due to the hardware noise, the actual ansatz and the resulting \rev{noisy} quantum state \rev{$\Rho$} differ from the ideal one\rev{s}. %

\rev{
The quantum noise $\mathcal{N}$ for a quantum system with state $\Rho$ is characterized as 
\begin{equation}
    \mathcal{N}(\Rho) = \sum_i \mat{E}_i \Rho \mat{E}_i^{\dagger},
\end{equation}
where $\mat{E}_i$ are Kraus operators satisfying $\sum_i \mat{E}_i \mat{E}_i^{\dagger} = \mat{I}$.
In this paper, we use amplitude and phase damping channels as the noise model because such a noise model has been shown to shift the optimal parameter of a VQA~\cite{sharma2020noise}. The amplitude damping noise describes the energy dissipation of quantum systems, whose Kraus operator formulation on a single qubit is 
\begin{equation}
    \mathcal{N}(\Rho) = \mat{E}_0 \Rho \mat{E}_0^\dagger + \mat{E}_1 \Rho \mat{E}_1^\dagger, 
\end{equation}
with $\mat{E}_0=\begin{pmatrix}1 & 0 \\
                         0 &\sqrt{1-p_{ad}} \end{pmatrix}$, $\mat{E}_1=\begin{pmatrix}0 & \sqrt{p_{ad}} \\
                         0 &0 \end{pmatrix}.$
The phase damping noise describes the quantum information loss without the energy loss, whose Kraus operator formulation on a single qubit is
\begin{equation}
    \mathcal{N}(\Rho) = \mat{E}_0 \Rho \mat{E}_0^\dagger + \mat{E}_1 \Rho \mat{E}_1^\dagger, 
\end{equation}
with $\mat{E}_0=\begin{pmatrix}1 & 0 \\
                         0 &\sqrt{1-p_{pd}} \end{pmatrix}$, $\mat{E}_1=\begin{pmatrix}0 & 0 \\
                         0 &\sqrt{p_{pd}} \end{pmatrix}.$
To integrate these two amplitude and phase damping noise channels, the combined Kraus operator is as follows 
\begin{equation}\label{eq:kraus_combine}
    \mathcal{N}(\Rho) = \sum_{i=0}^{2} \mat{E}_i \Rho \mat{E}_i^\dagger, 
\end{equation}
with $\mat{E}_0=\begin{pmatrix}1 & 0 \\
    0 &\sqrt{1-p_{ad}}\sqrt{1-p_{pd}} \end{pmatrix}$, 
    $\mat{E}_1=\begin{pmatrix}0 & \sqrt{p_{ad}} \\
    0 & 0 \end{pmatrix}$,
    $\mat{E}_2 = \begin{pmatrix}0 & 0\\
    0 &\sqrt{1-p_{ad}}\sqrt{p_{pd}} \end{pmatrix}$.
The parameters $p_{ad}$ and $p_{pd}$ are strongly related to the $T_1$ and $T_2$ time of quantum hardware. In this paper, we assume an equal probability of two damping channels $p_{ad} = p_{pd} = p$ for simplicity. We do not expect such noise modeling to capture practical hardware noise accurately but only use it for proof of concept.}

\subsection{Distributionally robust optimization formulation of VQAs}\label{sec:method}
\rev{We assume that we have access to the fixed noise model, i.e., the fixed Kraus presentation~\eqref{eq:kraus_combine}, but do not know the precise noise level $p$. We model $p$ as a random variable, which follows a certain PDF $\xi\sim \rho(\xi)$.} 
Let \rev{$f(\boldsymbol{\theta},\xi)=\mathrm{Tr}\left[\Rho(\boldsymbol{\theta},\xi) \mat{O} \right]$}
be the \rev{quantity of interest} evaluated by an ansatz parametrized by $\boldsymbol{\theta}$ under a noise level $\xi$\rev{, where $\Rho(\boldsymbol{\theta},\xi)$ is the resulting noisy quantum state}. A standard \rev{parameter optimization of a VQA} becomes stochastic programming:
\begin{equation}\label{eq:noise_aware_vqa}
    \min_{\boldsymbol{\theta}} \quad \mathbb{E}_{\rho(\xi)}[f(\boldsymbol{\theta},\xi)].
\end{equation}
\rev{Note that we here consider the one-dimensional noise level for simplicity. It can be seamlessly extended to a high-dimensional case.} 

However, due to the real-time \rev{fluctuation of quantum noise}, the \rev{actual} PDF of the noise level can shift and become unknown. As a result, we assume that $\rho(\xi)$ is not exactly known and it can be any PDF inside a set $\mathcal{P}$, which makes it impossible to obtain a deterministic value of $\mathbb{E}_{\rho(\xi)}[f(\boldsymbol{\theta},\xi)]$. As a result, we try to optimize the worst-case value of $ \mathbb{E}_{\rho(\xi)}[f(\boldsymbol{\theta},\xi)]$ by solving
\begin{equation}\label{eq:shift_aware_minmax}
    \min_{\boldsymbol{\theta}} \sup_{\rho(\xi) \in \mathcal{P}} \quad \mathbb{E}_{\rho(\xi)}[f(\boldsymbol{\theta},\xi)].
\end{equation}
When the uncertainty set degenerates to $\mathcal{P} = \{\rho(\xi) \}$, problem~\eqref{eq:shift_aware_minmax} degenerates to the standard stochastic optimization problem in ~\eqref{eq:noise_aware_vqa}. On the other hand, the problem degenerates to a robust optimization under the worst noise level when the PDF of a noise level degenerates to a Dirac function. 

The distributionally robust circuit optimization~\eqref{eq:shift_aware_minmax} may be intractable in practice because (a) $\mathcal{P}$ may contain an infinite number of PDFs describing process variations; (b) the min-max problem is hard to solve by nature; (c) we do not have an analytical form for $f(\boldsymbol{\theta},\xi)$ under the presence of noise. 

tion{The Proposed Solver} 
In this section, we properly define the PDF uncertainty set $\mathcal{P}$ and solve problem~\eqref{eq:shift_aware_minmax} leveraging distributionally robust Bayesian optimization (DRBO)~\cite{kirschner2020distributionally,nguyen2020distributionally,husain2022distributionally} developed recently in the machine learning community.

\subsection{Distribution Uncertainty Set}
We model the PDF uncertainty set $\mathcal{P}$ as a ball whose center is the nominal distribution ${\rho_0}(\xi)$ of the noise level and radius $\varepsilon$ is measured by a distribution divergence $\mathcal{D}$:
\begin{equation}\label{eq:def_ball}
    \mathcal{P}:= \mathcal{B}(\rho_0) =\{\rho: \mathcal{D}({\rho_0}, \rho) \leq \varepsilon \}.
\end{equation}
There are many options for the divergence $\mathcal{D}$, including Maximum Mean Discrepancy, Wasserstein distance, $\varphi$-divergence, etc~\cite{rahimian2019distributionally}. Here, we choose the Maximum Mean Discrepancy (MMD). 

MMD aims to compare the means of samples drawn from two distributions in a high-dimensional reproducing kernel Hilbert space (RKHS) induced by a positive definite kernel function~\cite{muandet2017kernel}. For the tractability of the problem, we discretize the noise level in a finite space $\Xi$ with $n$ parts. Then, let $\mathcal{H}_M$ be an RKHS with corresponding kernel $k_M: \Xi \times \Xi \rightarrow \R$, we can embed the distributions $\rho_0$ (similarly for $\rho$) into $\mathcal{H}_M$ via the mean embedding:
\begin{align*}
    m_{\rho_0} := \E_{\xi \sim \rho_0}[k_M(\xi,\cdot)], \quad \text{such that} \\
\langle  m_{\rho_0}, k_M(\xi^\prime,\cdot) \rangle = \E_{\xi \sim \rho_0}[k_M(\xi^\prime,\xi)], \forall \xi \in \Xi. 
\end{align*}

Then the MMD between two distributions $\rho_0$ and $\rho$ over $\Xi$ is defined as 
\begin{equation}\label{eq:embed_MMD}
    \mathcal{D}(\rho_0,\rho) := \|m_{\rho_0} - m_{\rho} \|_\mathcal{H},
\end{equation}
where $\|\cdot\|_\mathcal{H} = \sqrt{\langle \cdot, \cdot \rangle}$ is the Hilbert norm.  
Let $w_i = \rho_0{(\xi_i)}$ and $w^\prime_i = \rho{(\xi_i)}$ be the density probability of two discrete distributions, if we replace the expectation with the empirical expectation, i.e., $m_\rho = \sum_{i=1}^n w_i k_M(\xi_i,\cdot)$ and $m_{\rho^\prime} = \sum_{i=1}^n w_i^\prime w_i^\prime k_M(\xi_i,\cdot)$, Eq.~\eqref{eq:embed_MMD} can be written as:
\begin{equation}
    \mathcal{D}(\rho_0,\rho)=\sqrt{{(\mat{w}-\mat{w}^\prime)}^T \mat{M} {(\mat{w}-\mat{w}^\prime)}}, %
\end{equation}
where $M_{ij} = k_M(\xi_i,\xi_j)$ is the kernel matrix.  

\subsection{DRO Main Workflow} 
By modeling the distribution uncertainty set defined via MMD, the DRO problem~\eqref{eq:shift_aware_minmax} becomes tractable. The main steps are summarized below.
\begin{itemize}%
    \item {\bf Step 1.} Characterize the nominal noise distribution $\rho_0$. 
    \item {\bf Step 2.} Given a current $\boldsymbol{\theta}$, solve the inner problem to determine the worst-case PDF of $\xi$:
    \begin{align}
       & \sup_{\rho(\xi): \mathcal{D}({\rho_0}, \rho) \leq \varepsilon} \quad \mathbb{E}_{\rho(\xi)}[f(\vect{\theta},\xi)] = \notag \\
       & \max_{\substack{\mat{w}^\prime:{\|\mat{w}^\prime\|}_1=1,\\
        0\leq w_j^\prime \leq 1, \forall j\in[n],\\
        \sqrt{{(\mat{w}-\mat{w}^\prime)}^T \mat{M} {(\mat{w}-\mat{w}^\prime)}}\leq \varepsilon}}
        \langle \mat{w}^\prime, f_{\vect{\theta}} \rangle, \label{eq:inner}
    \end{align}
    where $f_{\vect{\theta}} := f(\vect{\theta},\cdot) \in \R^n$ is the output with a given parameter $\vect{\theta}$.
    \item {\bf Step 3.} Solve the outer problem to update $\vect{\theta}$.
    \begin{equation}\label{eq:outer}
         \min_{\vect{\theta}} \quad  \langle \mat{w}^\prime, f_{\vect{\theta}} \rangle
    \end{equation}
    \item \textbf{Step 4.} If not converge, go back to Step 2.
\end{itemize}
Specifically, Step 2 can be solved analytically via convex programming as it is second-order cone programming with respect to the worst-case distribution $\mat{w}^\prime$. Step 3 can be solved via a numerical optimizer. However, one of the challenges in steps 2 and 3 is that we need to simulate multiple $f(\vect{\theta},\xi)$, which can be expensive in practice. To address the computational issue, we apply a Bayesian optimization solver to the workflow. The key idea is to sequentially learn a surrogate model of $f(\vect{\theta},\xi)$ and optimize it by iteratively adding informative samples.  

\begin{algorithm}[H]
\caption{Overall DRBO algorithm with GP}\label{alg:DRBO_alg}
\begin{algorithmic}[1]
\Require Initial sample set ${\mathcal{S}_{0}=\{\vect{\theta}^i,\xi^i,f(\vect{\theta}^i,\xi^i)\}_{i=1}^M}$, reference PDF of noise level $\rho_0(\xi)$ with $\rho_0(\xi_i) = w_i, \forall i = [n]$, uncertainty ball radius $\varepsilon$, maximum iteration $T$
\Ensure The optimal circuit design variables $\vect{\theta}^\star$
\For{$t = 1,2,...,T$}
\State Construct a GP model as the probabilistic surrogate model $\hat{f}(\vect{\theta},\xi) = \mathcal{GP}(\vect{\theta},\xi)$ based on $\mathcal{S}_{t-1}$
\State Define $\text{LCB}(\vect{\theta},\xi):= \mu(\mathcal{GP}(\vect{\theta},\xi)) - \beta \cdot \sigma(\mathcal{GP}(\vect{\theta},\xi))$ 
\State Define the PDF of the worst-case distribution $\mat{w}^\prime := \argmax_{\mat{w}} \langle \mat{w}^\prime, \text{LCB}(\vect{\theta},\xi) \rangle$ s.t. ${\mat{w}^\prime:{\|\mat{w}^\prime\|}_1=1}$, ${0\leq w_j^\prime \leq 1, \forall j\in[n]}$, and ${{\|\mat{w}^\prime - \mat{w}\|}_{M} \leq \varepsilon}$
\State Solve the robust parameter $\vect{\theta}_{t} = \argmin_{\vect{\theta}} \langle \mat{w}^\prime, \text{LCB}(\vect{\theta},\xi) \rangle$ 
\State {Sample $K$ noise levels from the reference PDF $\xi_k \sim \rho_0$ and simulate $f(\vect{\theta}_t, \xi_k)$, for $k = 1, 2, \ldots, K$}
\State {Augment data set $\mathcal{S}_{t} \leftarrow {\mathcal{S}_{t-1} \cup \{(\vect{\theta}_t, \xi_t, f(\vect{\theta}_t, \xi_k)})\}_{k=1}^K$}
\EndFor 
\State {Return optimal $\vect{\theta}^\star$}
\end{algorithmic}
\end{algorithm}

\subsection{BO solver for DRO problem} 
Next, we explain how to solve DRO via Bayesian optimization with a few quantum circuit simulations. Bayesian optimization sequentially builds a probabilistic surrogate model of $f(\vect{\theta},\xi)$ and explores the design space by minimizing an acquisition function. The overall DRBO algorithm is summarized in Algorithm~\ref{alg:DRBO_alg}.

We first construct a probabilistic surrogate model $\hat{f}(\boldsymbol{\theta},\xi)$, which can estimate both the output and its uncertainty given an input $(\vect{\theta}, \xi)$. Here, we use the Gaussian process regression model $\mathcal{GP}(\boldsymbol{\theta},\xi)$ as the surrogate $\hat{f}(\boldsymbol{\theta},\xi)$. Then we use its lower confidence bound (LCB) to replace the original objective function $f(\boldsymbol{\theta},\xi)$ in Eqs.~\eqref{eq:inner} and~\eqref{eq:outer} in steps 2-4
\begin{equation}\label{eq:drbo_lcb}
    f(\vect{\theta},\xi) \rightarrow \text{LCB}(\boldsymbol{\theta},\xi) = \mu(\mathcal{GP}(\boldsymbol{\theta},\xi)) - \beta \cdot \sigma(\mathcal{GP}(\boldsymbol{\theta},\xi)),
\end{equation}
where $\mu(\cdot)$ and $\sigma(\cdot)$ denote the estimated mean and standard deviation, and $\beta$ is a parameter to balance the model exploitation and exploration. 

\textbf{Gaussian process surrogate.}
To build the Gaussian process regression (GPR) model, we need to predefine the mean function $m(\cdot)$ and the kernel function $k_{\mathcal{GP}}(\cdot,\cdot)$. Given a dataset $\mat{X} = \{\mat{x}^i\}_{i=1}^M = \{\vect{\theta}^i,\xi^i\}_{i=1}^M$ and their simulation outputs $\mat{y}=\{f({\mat{x}}^i)+\epsilon\}_{i=1}^M$, the GP model assumes that the simulation outputs follows a Gaussian distribution~\cite{rasmussen2003gaussian}: 
\begin{equation}
\text{Prob}(\mat{y}) = \mathcal{N}(\mat{y} \vert \vect{\mu}, \mat{K}),    
\end{equation} 
where $\vect{\mu} \in \R^M$ is the mean vector with $\mu_i = m(\mat{x})$, $\mat{K} \in \R^{M \times M}$ is the covariance matrix with $K_{i,j} = k_{\mathcal{GP}}(\mat{x}_i, \mat{x}_j)$. 
The measurement noise is characterized as a white noise $\epsilon$ in the simulation output.

Then, the GP model can offer a probabilistic prediction of a new data $\mat{x}^\prime$, $\mathcal{GP}(\mat{x}^\prime) \sim N(\mu({\mat{x}^\prime}),\sigma^2({\mat{x}^\prime}))$, as follows:
\begin{align*}
&\mu({\mat{x}^\prime}) = {\mat{k}_{GP}(\mat{x}^\prime, \mat{X})}^T{(\mat{K}+\epsilon^2 \mat{I})}^{-1} \mat{y} \\
&\sigma^2({\mat{x}^\prime}) = k_{\mathcal{GP}}(\mat{x}^\prime,\mat{x}^\prime) - {\mat{k}_{\mathcal{GP}}(\mat{x}^\prime,\mat{X})}^T{(\mat{K}+\epsilon^2 \mat{I})}^{-1}\mat{k}_{\mathcal{GP}}(\mat{X}, \vect{\theta}^\prime)    
\end{align*}
In our cases, we choose the prior mean as $m(\mat{x})=0$ and use the RBF kernel $k_{\mathcal{GP}}(\mat{x}_i,\mat{x}_j) = e^{-\frac{{\|\mat{x}_i - \mat{x})j\|}^2}{2 l^2}} $.
Note that \rev{the} kernel $k_{\mathcal{GP}}$ \rev{used} in GPR \rev{differs} from the kernel function $k_M$ \rev{employed} in MMD. \rev{Apart from} the Gaussian process, \rev{there are various other surrogate models that can be utilized.}

\textbf{Optimal selection.}
Regarding the selection of an optimal solution, it turns out to be non-trivial.  Since we want to estimate the expectation over the noise distribution, it will be too expensive to estimate with real quantum devices.  Instead, we choose the solution with maximized model posterior, i.e., we choose the $\vect{\theta} = \argmax_{\vect{\theta}} \E_{\rho(\mat{\xi})}[\mu(\hat{f}(\vect{\theta},\xi))] = \langle \mat{w}, \mu(\hat{f}_{\vect{\theta}}) \rangle$, where $\hat{f}_{\vect{\theta}}$ is the mean prediction from $\hat{f}$ with a given parameter $\vect{\theta}$. It is a common strategy for similar conditions~\cite {frazier2018tutorial,cakmak2020bayesian}.

In addition, an accurate estimation over $\hat{f}(\vect{\theta},\cdot)$ will benefit our model output. For this motivation, at the end of each iteration in Algorithm~\ref{alg:DRBO_alg}, we can add a batch of samples of $\xi$ from $\Xi$ in step 3. 
It can help build a better probabilistic model and fasten the BO solver convergence. 

\textbf{Remarks.}
One possible further improvement is to treat the BO solver as a warm-start procedure. After returning a few high-quality solutions from BO, we can conduct the local numerical optimization by taking them as the initial. The local search step may introduce additional computational cost and need more calling of $f(\vect{\theta},\xi)$ instead of the surrogate model, but it can lead to a potentially better solution. The hybrid of different solvers is also a common strategy in VQA parameter optimization~\cite{muller2022accelerating,lavrijsen2020classical}. 

The proposed distributionally robust optimization can easily degenerate into stochastic optimization or robust optimization. Stochastic optimization, namely Eq.~\eqref{eq:noise_aware_vqa}, does not consider the real-time change of the noise. Robust optimization degenerates the PDF of the noise level $\rho(\xi)$ to a single scalar. This case can easily lead to over-conservative parameter optimization.

The current distribution uncertainty set modeling of Maximum Mean Discrepancy has the great power of capturing the worst-case distribution. However, we need to discretize the noise level PDF in order to estimate the MMD efficiently. There exist some other approaches to modeling the uncertainty set that do not discretize the noise level, such as $f$-divergence modeling~\cite{husain2022distributionally}. Meanwhile, some alternative uncertainty set modeling could potentially reduce the computational overhead of estimating worst-case distribution~\eqref{eq:inner}~\cite{husain2022distributionally, farokhi2023distributionally, mohajerin2018data} iteratively. However, they may not perform well in our experiments due to the unfitted modeling of shifted noise level distribution. 
\begin{figure*}[t]
    \centering
    \includegraphics[width=\linewidth]{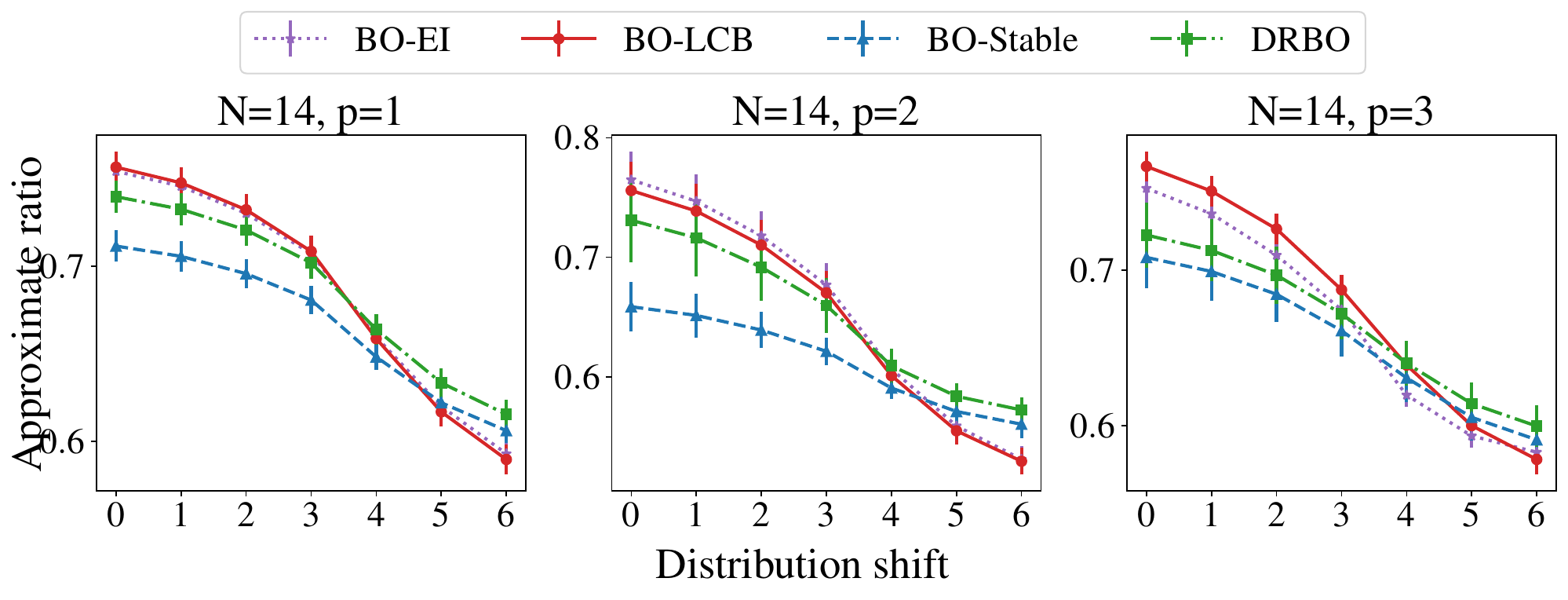}
    \caption{
    The results for solving $N=14$ $3$-regular graph MaxCut problems via QAOA. The $x$ axis denotes the significance of noise shift, where the noise level PDF is the reference one at $x=0$. The $y$ axis is the expectation of the approximation ratio of the QAOA solution evaluated at different noise \rev{PDFs}. We report the average result over $10$ non-isomorphic graphs. As we can see, the standard \rev{BO-LCB and BO-EI solutions have} the best performance under the reference noise. However, under an increasingly shifted noise, the DRBO solution begins to outperform the \rev{standard BO solutions}. Meanwhile, the BO-Stable solution is over-conservative with respect to the noise. It significantly scarifies the performance under the reference PDF and the slightly shifted PDFs to gain an improvement under significant shifts. These observations are consistent in the experiments with different QAOA depths.}
    \label{fig:maxcut_results}
\end{figure*}

\section{Numerical Experiments}\label{sec:result}
We validate the distributionally robust formulation of optimizing VQA parameters~\eqref{eq:shift_aware_minmax} in two widely used VQA applications: one is using QAOA for MaxCut and the other one is using VQE for a one-dimensional Heisenberg model.

Here, we conduct the numerical experiments on a simulator in order to adjust the noise level easily and correspondingly to validate the method. To apply the distributional robustness formulation in hardware experiments, the estimation of the noise model and noise level is another challenge, which is out of the scope of this work. 

\textbf{Baseline\rev{s}.} We compared the proposed DRBO solver to \rev{two} standard Bayesian optimization methods for solving stochastic optimization, \rev{one is with a lower confidence bound acquisition function (BO-LCB) and the other one is with an expectation improvement acquisition function (BO-EI)}~\cite{srinivas2009gaussian}, and robust Bayesian optimization (BO-Stable)~\cite{bogunovic2018adversarially}. 

In BO-LCB, we target problem~\ref{eq:noise_aware_vqa} with a fixed reference distribution of noise level $\rho_0(\xi)$ using a Bayesian optimization approach. We use the same GP surrogate model and its LCB as Eq.~\eqref{eq:drbo_lcb}, but without solving the outer problem~\eqref{eq:outer}. Specifically, the lines 4 and 5 of Algorithm~\ref{alg:DRBO_alg} are combined as solving ${\min_{\vect{\theta}} \langle \mat{w}, \text{LCB}(\vect{\theta},\xi) \rangle}$.

In BO-EI, we replace the above LCB function with the EI acquisition function. Let $\boldsymbol{\theta}^-$ be the best sample with the smallest value $f(\boldsymbol{\theta}^-, \xi)$ so far. The EI acquisition function is defined as:
\begin{equation}
\small{
    \text{EI}(\boldsymbol{\theta}, \xi) = \Phi(z) \left( f(\boldsymbol{\theta}^-, \xi) - \mu(\mathcal{GP}(\boldsymbol{\theta},\xi)) \right) + \phi(z) \sigma(\mathcal{GP}(\boldsymbol{\theta},\xi)),}
\end{equation}
where $z = \begin{cases}
    \frac{f(\boldsymbol{\theta}^-, \xi) - \mu(\mathcal{GP}(\boldsymbol{\theta},\xi))}{\sigma(\mathcal{GP}(\boldsymbol{\theta},\xi))} & \sigma(\mathcal{GP}(\boldsymbol{\theta},\xi)) > 0 \\
    0 & \sigma(\mathcal{GP}(\boldsymbol{\theta},\xi)) = 0
\end{cases}$, $\Phi(\cdot)$ and $\phi(\cdot)$ are the cumulative distribution function and the probability density function of the standard normal distribution. In the literature of applying BO for learning VQAs, there exists the usage of variant kernel functions and acquisition functions under different noisy environments. However, they can all be categorized as standard BO as they consider a fixed noise model only.

In BO-Stable, we target a shift-aware problem but only focus on the worst noise level instead of the worst distributional noise level. We use the same GP surrogate model and its LCB as Eq.~\eqref{eq:drbo_lcb}. Differing from using DRBO for problem~\eqref{eq:shift_aware_minmax}, the lines 4 and 5 of Algorithm~\ref{alg:DRBO_alg} are replaced with solving $\min_{\vect{\theta}} \text{LCB}(\vect{\theta}, \xi^\star)$ where given a $\vect{\theta}$, the worst $\xi^\star$ is defined as ${\xi^\star := \argmax_{\xi} \text{LCB}(\vect{\theta},\xi)}$.

To compare these different VQA parameter optimization methods, we first obtain their different optimized parameters and evaluate them under different levels of noise $\rho(\xi)$.   
\rev{In the simulation, we used the $\mathsf{qiskit\_aer}$ noisy simulator with the statevector backend, whose simulation algorithm is the Monte Carlo trajectory approach. In one circuit trajectory, a Kraus operator is randomly applied on an ideal gate with probability that is defined by the noise channel. The multi-qubit noise channel is defined as the tensor product of single-qubit noise ones~\eqref{eq:kraus_combine}.}

\subsection{\rev{Numerical} Experiments on QAOA}
\begin{figure*}[t]
    \centering
    \includegraphics[width=\linewidth]{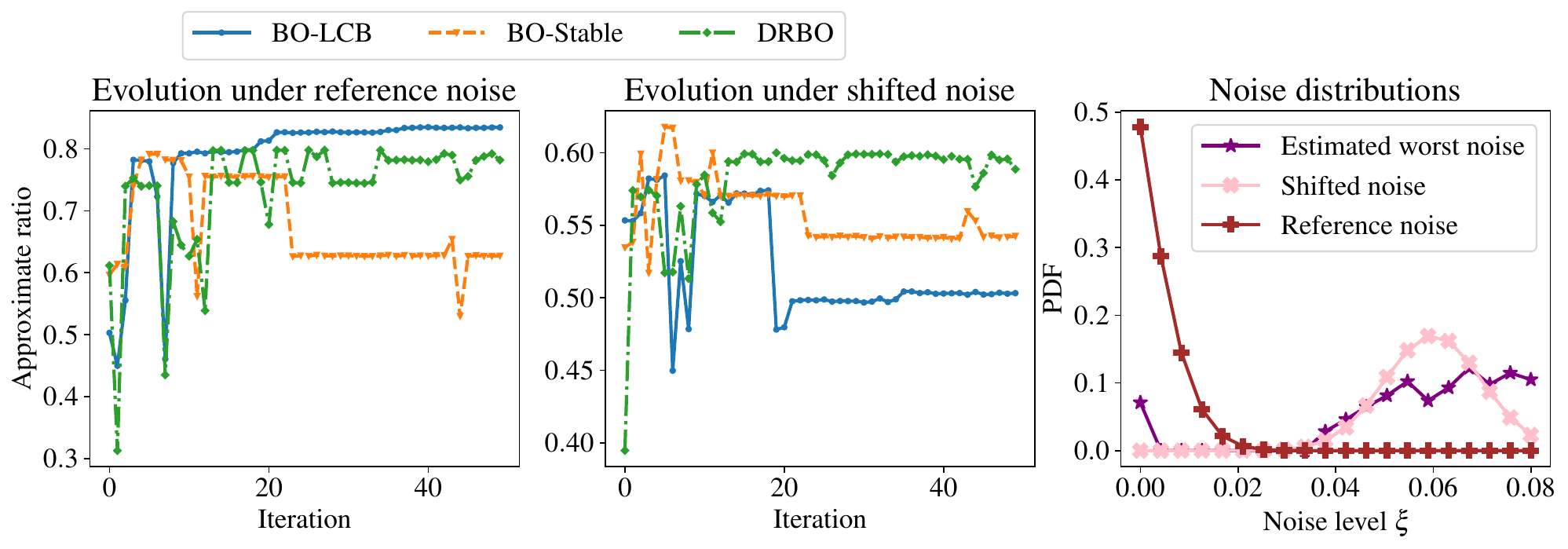}
    \caption{One example of evolving of solution in different BO algorithms. The $x$ axis is the iterations in a BO algorithm, and the $y$ axis is the expectation of cost function evaluated over noise level at a $\vect{\theta}$. The evaluated $\vect{\theta}$ at one iteration is obtained by maximizing the model posterior, which is unnecessary to be the explored $\vect{\theta}$ at that iteration. Under the reference noise PDF, the BO-LCB algorithm converges to a better solution, while the DRBO converges to a better solution under the shifted noise. The rightmost figure shows the example PDFs of the reference noise level, shifted noise level, and the estimated worst-case noise level from the DRBO algorithm.}
    \label{fig:sample_evolution}
\end{figure*}

\begin{figure}[t]
    \centering
    \includegraphics[width=0.7\linewidth]{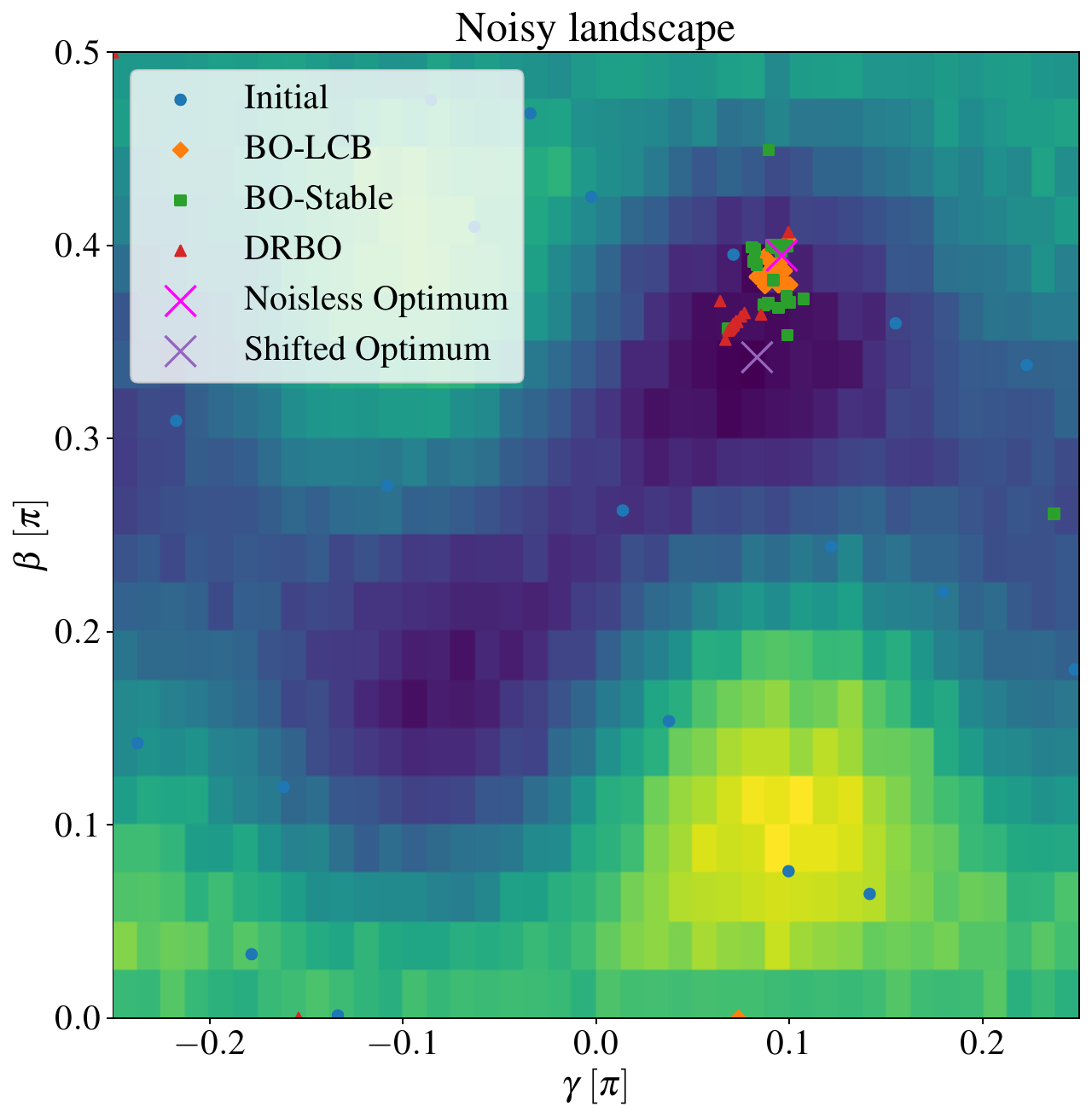}
    \caption{An example of explored $\boldsymbol{\theta}$ in $p=1$ noisy QAOA cost landscape in a $N=8$ MaxCut problem. The optimum $\boldsymbol{\theta}$ differs from the noiseless optimum. Compared to the BO-LCB and BO-Stable, the DRBO explores the parameter space that performs well under the shifted noise.}
    \label{fig:sample_in_landscape}
\end{figure}
\begin{figure*}[t]
    \centering
    \includegraphics[width=\linewidth]{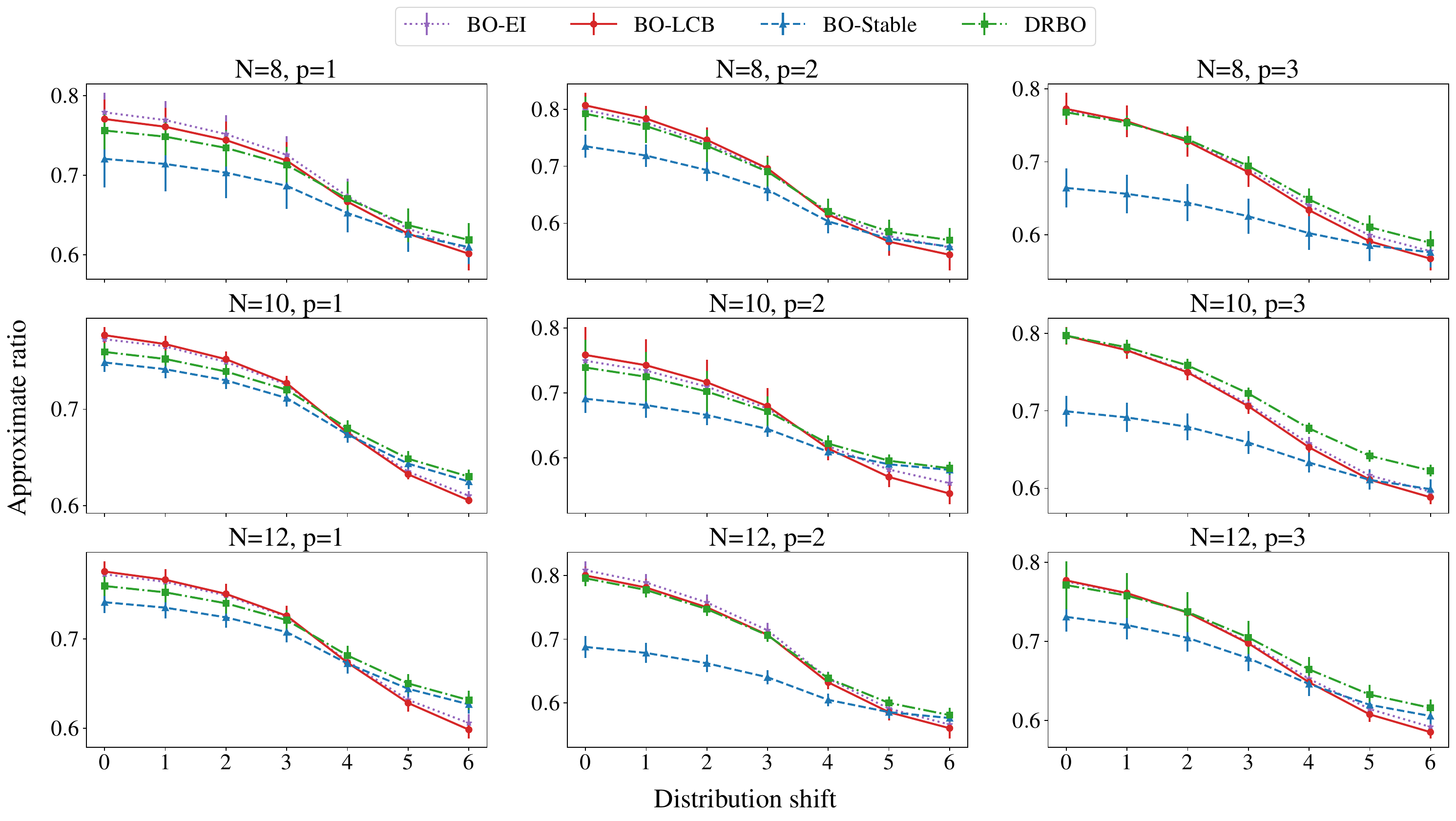}
    \caption{More results on the Max-Cut experiments with graph sizes $N=8,10,12$ and QAOA depth $p=1,2,3$.  While potentially sacrificing the performance under the reference noise a little, the DRBO solution performs better \rev{than standard BO methods} under the significantly shifted noise. Meanwhile, BO-Stable solutions are over-conservative.}
    \label{fig:more_maxcut_results}
\end{figure*}
\begin{figure}[t]
    \centering
    \includegraphics[width=0.7\linewidth]{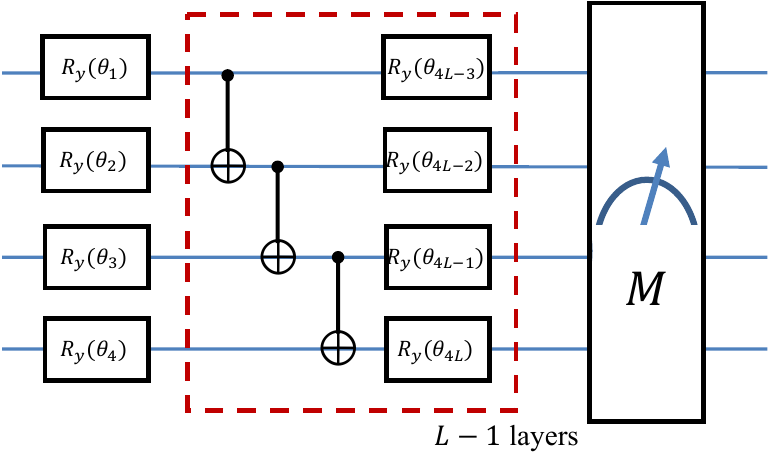}
    \caption{The schematic of the hardware-efficient ansatz for variational quantum eigensolver. The layer of two-qubit entanglement and one-qubit rotation gates are repeated for $L-1$ times.}
    \label{fig:hea_vqe}
\end{figure}

\begin{figure}[t]
    \centering
    \includegraphics[width=0.7\linewidth]{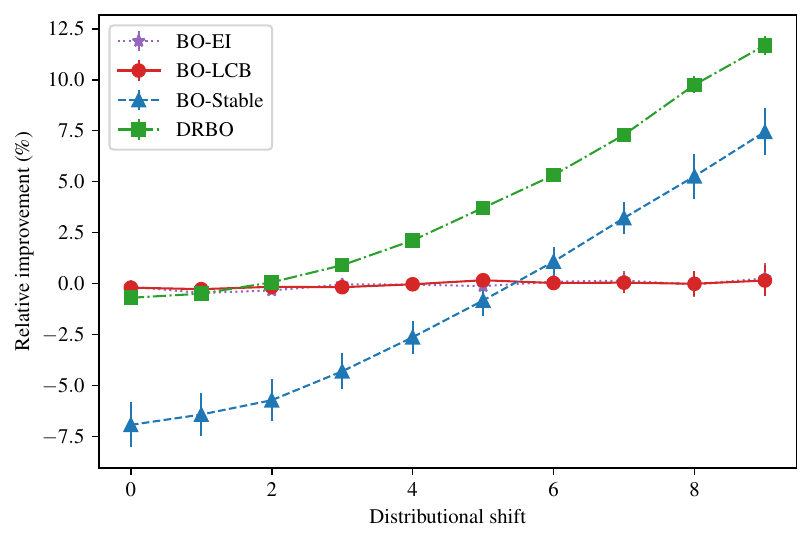}
    \caption{%
    The results for solving the ground energy of a $6$-spin, $J=1, B=0.2$ one-dimensional Heisenberg model via VQE with two-layer hardware efficient ansatz. The $x$ axis denotes the significance of noise shift, where the noise level PDF is the reference one at $x=0$. We first obtained the optimal parameter $\theta_0$ in a noiseless simulation, which solves the problem perfectly with $f(\vect{\theta}_0) = -4.8$. Then we report the relative improvement of the energy $\frac{ \mathbb{E}_{\rho(\xi)}[f(\vect{\theta}, \xi)] - \mathbb{E}_{\rho(\xi)}[f(\vect{\theta}_0, \xi)])}{\mathbb{E}_{\rho(\xi)}[f(\vect{\theta}_0, \xi)]}$. Under all the shifted distributions, BO-LCB performs close to $\vect{\theta}_0$. DRBO scarifies limited performance under mild noise and performs much better than the BO-LCB\rev{, BO-EI,} and noiseless optimal $\vect{\theta}_0$ in significantly shifted noise. \rev{BO-LCB and BO-EI are almost overlapping since they both have solutions close to $\vect{\theta}_0$.} While the BO-Stable can also find the robust parameter under the shifted noise, \rev{it does not perform as well as DRBO}, especially when the noise shift is mild. 
    }
    \label{fig:heisenberg_results_JB06}
\end{figure}

\begin{figure}[t]
    \centering
    \includegraphics[width=\linewidth]{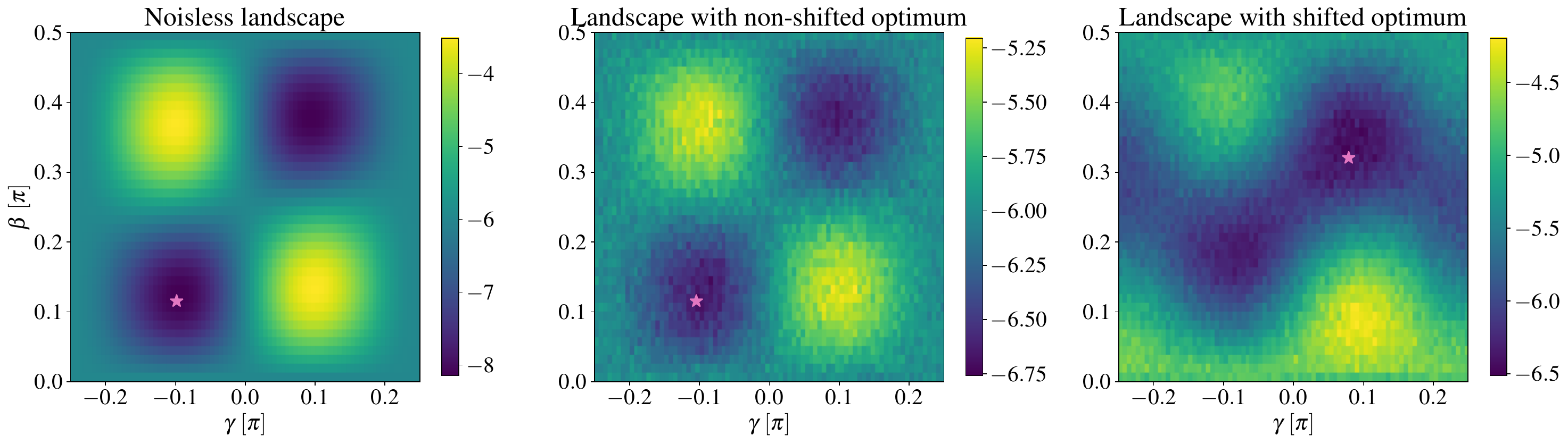}
    \caption{Example of energy landscapes under different noise models. The left heatmap is the depth-1 QAOA landscape for the MaxCut problem. The color denotes the solved energy. The optimal point in this landscape is highlighted as a triangle. The middle heatmap is the landscape under simple Pauli errors, which has been shown not to change the VQA optimal and uniformly flatten the landscape. The right heatmap is the landscape under the phase and amplitude damping noise, where the optimum is shifted and the energy landscape has a different shape. Under the noise, both the middle and the right ones have worse energies than the left noise-free landscape.}
    \label{fig:example_landscape}
\end{figure}

QAOA is a leading variational quantum algorithm for combinatorial optimization problems. It alternatively applies two operators, a phase-separation operator and a mixer operator, to drive a quantum system to the target solution state. \rev{A noiseless} QAOA solution is denoted as $\vect{\psi}(\vect{\theta}) = e^{-i\beta_p \mat{H}_M} e^{-i\gamma_p \mat{H}_P} \cdots e^{-i\beta_1 \mat{H}_M} e^{-i\gamma_1 \mat{H}_P} \vert \psi_0 \rangle$.

We will take the MaxCut problem as a case study of QAOA. Given a graph $G=(V,E)$ with vertices $V$ and edges $E$, the MaxCut problem aims to find a cut that partitions the graph vertices into two sets with the largest number of edges. Its cost function is written as 
\begin{equation}
    C = \sum_{(i,j) \in E} 1-s_i s_j,
\end{equation}
where $s_i$ and $s_j$ are binary variables associated to the vertices in $V$, which assume value $1$ or $-1$ depending on which of the two partitions defined by the cut are assigned. Its cost Hamiltonian is defined as $\mat{H}_C = \sum_{(u,v)\in E} \frac{1}{2}(\mat{I}-\mat{Z}_i\mat{Z}_j)$, where $\mat{Z}_i$ denotes a Pauli-$\zgate$ operator. 

In applying QAOA for solving the MaxCut problem, given a noise model with noise level $\xi$, we aim to optimize the QAOA parameters $\vect{\theta}=(\vect{\gamma},\vect{\beta})$ such that the resulting \rev{quantum state $\Rho(\boldsymbol{\theta},\xi)$} has minimal energy \rev{$f(\vect{\theta},\xi)=\Tr\left[\Rho(\boldsymbol{\theta},\xi) \mat{H}_C\right]$}
Considering the uncertainty and \rev{fluctuation} of noise level, we aim to find parameters $\vect{\theta}$ that make QAOA performance more robust towards the \rev{shifted} noise by solving the DRO problem~\eqref{eq:shift_aware_minmax}. 

Here, we discretize the noise level into 20 bins in $[0,0.08]$ evenly. We assume the reference noise follows a truncated Gaussian distribution, with the \rev{noise fluctuation shifting} its mean to a larger value. We first generate a truncated Gaussian distribution with mean $-0.01$ and standard deviation $0.01$. We estimate the probability density at each discretized level and do the normalization to obtain the reference PDF of the noise level. We follow a similar procedure to generate the PDF of shifted noise by shifting the mean of the initial truncated Gaussian distribution.

To begin with, for a depth-$p$ QAOA ansatz, we initialize the sampling set ${\mathcal{S}_{0}=\{\vect{\theta}^i,\xi^i,f(\vect{\theta}^i,\xi^i)\}_{i=1}^M}$ by taking $M=20p$, where ${\vect{\theta}}^i$ is drawn from the design space based on a Latin hypercube approach~\cite{ye1998orthogonal}, and the noise level samples ${\xi}^i$ are drawn from the reference distribution $\rho_0(\xi)$. We set the maximum BO iterations as $T=20p$. 

As shown in Fig.~\ref{fig:maxcut_results}, we evaluate different BO-based parameter optimization results on $10$ graphs with degree-$3$ and graph size $N=14$. We report the average approximation ratio results under different shifted noise levels. The $x$-axis denotes the index of the levels of noise shift, with a higher one denoting a more significant shift, and index-$0$ denotes the reference noise. Since we solve the optimal $\vect{\theta}$ under shifted noise, the DRBO-solved QAOA is expected to perform worse than the one solved from a standard BO solver under the reference noise. However, as the noise shift becomes more and more significant, the DRBO solution begins to show its advantages. Notably, \rev{BO-Stable performs better than BO-LCB and BO-EI under a significantly shifted noise as well.} However, it is also over-conservative under the reference noise since it only considers a single worst noise level. The results and observations are consistent over different QAOA depths.

We plot the solution during the BO iterations in Fig.~\ref{fig:sample_evolution}. During the iteration, the performance is evaluated under the optimal solution selected from the maximum posterior rather than from the solution of an acquisition function. We show the performance evaluated under the reference noise and the shifted noise. During the iterations, DRBO consistently converges to a shifted-noise preferred solution while the LCB converges to a reference-noise preferred one. We also show the PDFs of the reference noise, shifted noise, and the worst-case \rev{that is estimated by} the DRBO algorithm. We can see that the MMD approach successfully captures the shifted noise under the worse-case distribution, enabling the DRBO to explore the parameters space that performs better under shifted noise. 

More results on MaxCut with graph size $N=8,10,12$ and QAOA depth $p=1,2,3$ are shown in Fig.~\ref{fig:more_maxcut_results}. The results are consistent with the ones in Fig.~\ref{fig:maxcut_results}. The DRBO solution performs better than the baselines under significantly shifted noise, which demonstrates that our method could optimize the VQA parameters that are more robust to the \rev{shifted} noise. One example of the newly sampled $\vect{\theta}$ with $p=1$ is plotted in Fig.~\ref{fig:sample_in_landscape}. The DRBO algorithm explores the parameter space toward the optimal one under a shifted noise, while the other algorithms exploit the space surrounding the optimal parameter under the reference noise. Therefore, the DRBO could find a parameter that performs better under shifted noise.

\subsection{\rev{Numerical} Experiments on VQE}
Variational quantum eigensolver (VQE) is another popular VQA, specifically designed to simulate quantum systems and find the ground state energy of quantum systems. We use the VQE algorithm with a hardware-efficient ansatz~\cite{kandala2017hardware} for simulating the ground energy of a one-dimensional Heisenberg model defined as: 
\begin{equation}
    \mat{H} = J \sum_i \mat{X}_i\mat{X}_{i+1} + \mat{Y}_i\mat{Y}_{i+1} + \mat{Z}_i\mat{Z}_{i+1}
    + B \sum_i \mat{Z}_i,
\end{equation}
where $J$ is the strength of the spin–spin interaction and $B$ is the magnetic field along the $Z$ direction. Here, we use a hardware-efficient ansatz to implement the VQE algorithm. Given an ansatz parameterized by $\vect{\theta}$ and under a noise model with noise level $\xi$, we denote the state as %
\rev{$\Rho(\boldsymbol{\theta},\xi)$}. The \rev{noisy} VQE algorithm cost function is defined as \rev{$f(\vect{\theta},\xi)=\Tr\left[\Rho(\boldsymbol{\theta},\xi) \mat{H}\right]$}
Here, we aim to find the ansatz parameters that lead to robust performance under shifted noise by solving the DRO problem~\eqref{eq:shift_aware_minmax}.

Here, we discretize the noise level into $20$ bins in $[0,0.08]$ evenly. We assume the reference noise follows a truncated Gaussian distribution, and the \rev{changing noise shifts} its mean to a larger value. The hardware-efficiency ansatz is set up as shown in Fig.~\ref{fig:hea_vqe}. The number of parameters grows quickly and becomes challenging for a BO solver. For simplicity, we only optimize the last $N$ parameters and fix the others, similar to the idea of layer-wise optimization in Ref.~\cite{liu2022layer}. For the demonstration purpose, the fixed parameters are obtained through a multi-start classical optimization routine. We follow the same procedure as QAOA to set up the noise level distribution of both the reference and the shifted ones. 

In Fig.~\ref{fig:heisenberg_results_JB06}, we show the ground energy solved for a $6$-spin system with $J=1, B=0.2$, whose ground state is highly entangled, from a two-layer hardware-efficiency ansatz. Aiming to optimize the last layer parameters, we initialize the sampling set with $M=40$ and set the maximum BO iteration $T=40$ as well. Similar to the QAOA results, DRBO performs better than \rev{BO-LCB and BO-EI} under a significantly shifted noise. \rev{Furthermore, the DRBO solution minimally compromises the reference noise performance, whereas Bo-Stable tends to produce overly conservative results.}

\section{Discussion}\label{sec:discussion} 
\subsection{\rev{Related Works of Bayesian optimization}}
\rev{Bayesian optimization has been a prominent technique in addressing VQA learning tasks. For instance, \cite{self2021variational} implemented a parallel optimization scheme, enhancing the efficiency of optimizing VQA parameters across multiple similar problems by leveraging information sharing within Bayesian optimization. ~\cite{duffield2023bayesian} used Bayesian inference to reduce the redundancy of the parameterized circuit, ending with a shallower and more robust VQA ansatz. 
~\cite{iannelli2021noisy} used the standard Bayesian optimization techniques to optimize parameters for noisy VQE. ~\cite{muller2022accelerating} proposed a hybrid approach, utilizing Bayesian optimization solutions as warm starts and transitioning to multi-start local search from optimized Bayesian samples. ~\cite{tibaldi2023bayesian} specifically applied Bayesian optimization to optimize QAOA parameters. 
~\cite{finvzgar2024designing} demonstrated the efficiency of Bayesian optimization in optimizing parameterized quantum annealing schedules. 
~\cite{tamiya2022stochastic} introduced stochastic gradient line Bayesian optimization, leveraging Bayesian optimization to adjust step sizes in stochastic gradient descent, thereby reducing measurement-shot costs in optimizing VQA parameters.
~\cite{kim2023quantum} extended standard QAOA with two mixers and optimized circuit parameters through Bayesian optimization.
~\cite{benitez2024bayesian} utilized a VQE-kernel to construct a Gaussian process model, leveraging specific circuit properties as physics-informed priors and introducing a novel acquisition function to exploit the inductive bias of the kernel. 
~\cite{ravi2022cafqa} aimed to initialize good ansatz by fully exploring the Clifford parameter space through Bayesian optimization, where all simulations can be performed classically.
~\cite{cheng2024quantum} introduced a novel Bayesian optimization approach, optimizing QAOA parameters by constructing a surrogate model with constraints derived from two adaptive regions.}

\rev{Among all the above literature, it is hard to justify the best Bayesian optimization techniques since they use different surrogate modeling techniques, acquisition functions under different noise models, and target different applications. However, they can be all viewed as standard BO setups as they all aim to optimize either the parameter or ansatz architecture of VQA under a fixed noise environment. None of them shared the same shift-aware problem setup as us. }

\subsection{Landscape shift}
Ref.~\cite{sharma2020noise} has shown that the optimal variational parameters are unaffected by a broad class of noise models, such as measurement noise, gate noise, and Pauli channel noise. This phenomenon is called optimal parameter resilience. Meanwhile, some noise can shift the location of minima. A rich of work has studied how quantum noise can influence the VQA landscape~\cite{wang2021can,wang2021noise,fontana2022non}. We highlight that the shift location of optimal parameters motivates our work, i.e., given that the optimal parameter will change under different (shifted) noise, we aim to find \rev{optimal parameters} with robust performance under the shifted noise environment.

In our simulation, we use the phase and amplitude damping noise model, which has been shown to change the values of optimal parameters. The landscape with a changed or unchanged optimum is illustrated in Fig.~\ref{fig:example_landscape}. 

\subsection{Variant problem formulation}\label{sec:time_varying}
\textbf{Radius varying formulation.}
Beyond optimization under \rev{varying} noise, another case where the distributionally robust optimization can be applied is \rev{to calibrate the noise estimation.} \rev{Assume we} do not have a precise enough estimation \rev{of} the real noise level distribution as the reference distribution. \rev{The key idea is that} as we collect more data on the noise level, we can have a more accurate estimation of its PDF. Therefore, as the iteration continues, we can gradually refine the center $\rho_0(\xi)$ of the uncertainty ball and reduce its radius $\varepsilon$.

\textbf{Gate error modeling.}
Beyond modeling hardware noise, another possible modeling is on the gate error $f(\vect{\theta}+\vect{\xi})$ \rev{of a} parameterized quantum circuit, which assumes that the gate parameters are not exactly implemented but suffer from some \rev{coherent} errors. In such a formulation, under different error levels of $\vect{\xi}$, the optimal $\vect{\theta}$ will \rev{clearly} have different values. The DRO formulation can optimize the VQA to find parameters that are robust to the shifted gate errors. A robustness analysis of such a formulation is discussed in~\cite{rabinovich2022gate}. \rev{It is also of great interest to connect the gate error with a more detailed control error and apply the proposed shift-aware optimization at the physical level, but it is out of the scope of this manuscript.}

\section{Conclusion and opportunities}
\rev{Quantum} noise has been a major obstacle to the \rev{practical} applications of near-term quantum computers, \rev{particularly in} variational quantum algorithms (VQAs). \rev{Despite} various error mitigation techniques have been intensively studied, \rev{the dynamic nature of quantum noise presents a formidable challenge.} \rev{Optimized VQA parameters may perform suboptimally when exposed to different noise environments.}

In this paper, we have presented a distributionally robust optimization formulation \rev{designed to enhance the robustness of VQA parameters against varying quantum noise conditions}. Our approach leverages a distributionally robust Bayesian optimization (DRBO) solver, efficiently tackling the proposed formulation. We validate the proposed method in two \rev{widely recognized} VQA benchmarks: QAOA for MaxCut and VQE with hardware-efficient ansatz for \rev{an} one-dimensional Heisenberg model. The proposed distributionally robust optimization formulation does not \rev{aim to mitigate inherent quantum noise directly}. \rev{Instead,} it \rev{addresses} the noise at the \rev{algorithmic} level. It can be potentially integrated with various error mitigation techniques to further improve the VQA robustness. 

Our formulation can be more impactful \rev{in scenarios involving} large-size problems. \rev{As we scale up to larger problems or deeper VQA ansatz, even minor fluctuations in noise levels can significantly affect performance. For instance, implementing QAOA for larger problems necessitates deeper circuit depths, amplifying the influence of even slight shifts in noise levels on VQA performance}. Therefore, \rev{optimization VQA parameters under such shifts becomes increasingly critical.}

To integrate the proposed distributionally robust formulation into more practical use cases, a better knowledge of noise models is highly desired.
\rev{More specifically, in this paper, we characterize quantum noise by a fixed noise model, such as Eq.~\eqref{eq:kraus_combine}, and model the noise level as a random variable with an unknown PDF. Such mathematical modeling may not capture the actual hardware noise well since the noise characterization is in principle challenging. Another point is that we also need to model the uncertainty ball carefully. To make sure the optimized parameters perform well under the actual shifted noise, we need to tune the radius of uncertainty ball $\varepsilon$ such that the shifted noise is lying with the ball and is close to the worst case within the ball. Otherwise, the proposed DRBO may become over-conservative as shown in Figs.~\ref{fig:maxcut_results}~\ref{fig:more_maxcut_results}~\ref{fig:heisenberg_results_JB06}.}

To improve the \rev{solver of} distributionally robust optimization, some better techniques that do not need to discretize the noise level or efficiently handle high-dimensional parameter optimization \rev{of $\boldsymbol{\theta}$} can be developed. 

We also have applied a similar distributionally robust optimization formulation for classical circuit optimization~\cite{he_iccad2023}, where we identified the shifts of process variations. \rev{In this paper, we focus on handling the parameter optimization of noisy variation quantum algorithms. Differing from ~\cite{he_iccad2023}, the noise source and modeling in this paper are distinct, characterized by a fixed noise model with different levels of strength. We used MMD to define the uncertainty ball such that the shift-aware problem is solved in a two-step process. Additionally, we study its quantum applications from the energy landscape perspective, which is much less studied in classical circuit applications. There is a great thread studying the energy landscape of VQAs like Refs~\cite{hao2023enabling, perez2024analyzing, hao2024variational}. We believe our shift-aware optimization is interesting and could be inspiring to the community.}
\section*{Acknowledgements}
Z.H. and Z.Z. are supported in part by NSF CCF \# 1846476. B.P. is supported by the U.S. Department of Energy, Office of Science and National Quantum Information Science Research Centers, DOE Q-NEXT Center (Grant No. DOE 1F-60579). Y.A. is supported in part by funding from the Defense Advanced Research Projects Agency. This work used in part the resources of the Argonne Leadership Computing Facility, which is Department of Energy Office of Science User Facility supported under Contract DE-AC02-06CH11357. The views, opinions and/or findings expressed are those of the authors and should not be interpreted as representing the official views or policies of the Department of Defense, the Department of Energy, or the U.S. Government.

\bibliography{reference}

\begin{thebibliography}{78}%
\makeatletter
\providecommand \@ifxundefined [1]{%
 \@ifx{#1\undefined}
}%
\providecommand \@ifnum [1]{%
 \ifnum #1\expandafter \@firstoftwo
 \else \expandafter \@secondoftwo
 \fi
}%
\providecommand \@ifx [1]{%
 \ifx #1\expandafter \@firstoftwo
 \else \expandafter \@secondoftwo
 \fi
}%
\providecommand \natexlab [1]{#1}%
\providecommand \enquote  [1]{``#1''}%
\providecommand \bibnamefont  [1]{#1}%
\providecommand \bibfnamefont [1]{#1}%
\providecommand \citenamefont [1]{#1}%
\providecommand \href@noop [0]{\@secondoftwo}%
\providecommand \href [0]{\begingroup \@sanitize@url \@href}%
\providecommand \@href[1]{\@@startlink{#1}\@@href}%
\providecommand \@@href[1]{\endgroup#1\@@endlink}%
\providecommand \@sanitize@url [0]{\catcode `\\12\catcode `\$12\catcode `\&12\catcode `\#12\catcode `\^12\catcode `\_12\catcode `\%12\relax}%
\providecommand \@@startlink[1]{}%
\providecommand \@@endlink[0]{}%
\providecommand \url  [0]{\begingroup\@sanitize@url \@url }%
\providecommand \@url [1]{\endgroup\@href {#1}{\urlprefix }}%
\providecommand \urlprefix  [0]{URL }%
\providecommand \Eprint [0]{\href }%
\providecommand \doibase [0]{https://doi.org/}%
\providecommand \selectlanguage [0]{\@gobble}%
\providecommand \bibinfo  [0]{\@secondoftwo}%
\providecommand \bibfield  [0]{\@secondoftwo}%
\providecommand \translation [1]{[#1]}%
\providecommand \BibitemOpen [0]{}%
\providecommand \bibitemStop [0]{}%
\providecommand \bibitemNoStop [0]{.\EOS\space}%
\providecommand \EOS [0]{\spacefactor3000\relax}%
\providecommand \BibitemShut  [1]{\csname bibitem#1\endcsname}%
\let\auto@bib@innerbib\@empty
\bibitem [{\citenamefont {Cerezo}\ \emph {et~al.}(2021)\citenamefont {Cerezo}, \citenamefont {Arrasmith}, \citenamefont {Babbush}, \citenamefont {Benjamin}, \citenamefont {Endo}, \citenamefont {Fujii}, \citenamefont {McClean}, \citenamefont {Mitarai}, \citenamefont {Yuan}, \citenamefont {Cincio} \emph {et~al.}}]{cerezo2021variational}%
  \BibitemOpen
  \bibfield  {author} {\bibinfo {author} {\bibfnamefont {M.}~\bibnamefont {Cerezo}}, \bibinfo {author} {\bibfnamefont {A.}~\bibnamefont {Arrasmith}}, \bibinfo {author} {\bibfnamefont {R.}~\bibnamefont {Babbush}}, \bibinfo {author} {\bibfnamefont {S.~C.}\ \bibnamefont {Benjamin}}, \bibinfo {author} {\bibfnamefont {S.}~\bibnamefont {Endo}}, \bibinfo {author} {\bibfnamefont {K.}~\bibnamefont {Fujii}}, \bibinfo {author} {\bibfnamefont {J.~R.}\ \bibnamefont {McClean}}, \bibinfo {author} {\bibfnamefont {K.}~\bibnamefont {Mitarai}}, \bibinfo {author} {\bibfnamefont {X.}~\bibnamefont {Yuan}}, \bibinfo {author} {\bibfnamefont {L.}~\bibnamefont {Cincio}}, \emph {et~al.},\ }\bibfield  {title} {\bibinfo {title} {Variational quantum algorithms},\ }\href@noop {} {\bibfield  {journal} {\bibinfo  {journal} {Nature Reviews Physics}\ }\textbf {\bibinfo {volume} {3}},\ \bibinfo {pages} {625} (\bibinfo {year} {2021})}\BibitemShut {NoStop}%
\bibitem [{\citenamefont {Moll}\ \emph {et~al.}(2018)\citenamefont {Moll}, \citenamefont {Barkoutsos}, \citenamefont {Bishop}, \citenamefont {Chow}, \citenamefont {Cross}, \citenamefont {Egger}, \citenamefont {Filipp}, \citenamefont {Fuhrer}, \citenamefont {Gambetta}, \citenamefont {Ganzhorn}, \citenamefont {Kandala}, \citenamefont {Mezzacapo}, \citenamefont {Müller}, \citenamefont {Riess}, \citenamefont {Salis}, \citenamefont {Smolin}, \citenamefont {Tavernelli},\ and\ \citenamefont {Temme}}]{Moll_2018}%
  \BibitemOpen
  \bibfield  {author} {\bibinfo {author} {\bibfnamefont {N.}~\bibnamefont {Moll}}, \bibinfo {author} {\bibfnamefont {P.}~\bibnamefont {Barkoutsos}}, \bibinfo {author} {\bibfnamefont {L.~S.}\ \bibnamefont {Bishop}}, \bibinfo {author} {\bibfnamefont {J.~M.}\ \bibnamefont {Chow}}, \bibinfo {author} {\bibfnamefont {A.}~\bibnamefont {Cross}}, \bibinfo {author} {\bibfnamefont {D.~J.}\ \bibnamefont {Egger}}, \bibinfo {author} {\bibfnamefont {S.}~\bibnamefont {Filipp}}, \bibinfo {author} {\bibfnamefont {A.}~\bibnamefont {Fuhrer}}, \bibinfo {author} {\bibfnamefont {J.~M.}\ \bibnamefont {Gambetta}}, \bibinfo {author} {\bibfnamefont {M.}~\bibnamefont {Ganzhorn}}, \bibinfo {author} {\bibfnamefont {A.}~\bibnamefont {Kandala}}, \bibinfo {author} {\bibfnamefont {A.}~\bibnamefont {Mezzacapo}}, \bibinfo {author} {\bibfnamefont {P.}~\bibnamefont {Müller}}, \bibinfo {author} {\bibfnamefont {W.}~\bibnamefont {Riess}}, \bibinfo {author} {\bibfnamefont {G.}~\bibnamefont {Salis}}, \bibinfo {author} {\bibfnamefont {J.}~\bibnamefont
  {Smolin}}, \bibinfo {author} {\bibfnamefont {I.}~\bibnamefont {Tavernelli}},\ and\ \bibinfo {author} {\bibfnamefont {K.}~\bibnamefont {Temme}},\ }\bibfield  {title} {\bibinfo {title} {Quantum optimization using variational algorithms on near-term quantum devices},\ }\href@noop {} {\bibfield  {journal} {\bibinfo  {journal} {Quantum Science and Technology}\ }\textbf {\bibinfo {volume} {3}},\ \bibinfo {pages} {030503} (\bibinfo {year} {2018})}\BibitemShut {NoStop}%
\bibitem [{\citenamefont {Egger}\ \emph {et~al.}(2021)\citenamefont {Egger}, \citenamefont {Mare{\v{c}}ek},\ and\ \citenamefont {Woerner}}]{egger2021warm}%
  \BibitemOpen
  \bibfield  {author} {\bibinfo {author} {\bibfnamefont {D.~J.}\ \bibnamefont {Egger}}, \bibinfo {author} {\bibfnamefont {J.}~\bibnamefont {Mare{\v{c}}ek}},\ and\ \bibinfo {author} {\bibfnamefont {S.}~\bibnamefont {Woerner}},\ }\bibfield  {title} {\bibinfo {title} {Warm-starting quantum optimization},\ }\href@noop {} {\bibfield  {journal} {\bibinfo  {journal} {Quantum}\ }\textbf {\bibinfo {volume} {5}},\ \bibinfo {pages} {479} (\bibinfo {year} {2021})}\BibitemShut {NoStop}%
\bibitem [{\citenamefont {He}\ \emph {et~al.}(2023)\citenamefont {He}, \citenamefont {Shaydulin}, \citenamefont {Chakrabarti}, \citenamefont {Herman}, \citenamefont {Li}, \citenamefont {Sun},\ and\ \citenamefont {Pistoia}}]{he2023alignment}%
  \BibitemOpen
  \bibfield  {author} {\bibinfo {author} {\bibfnamefont {Z.}~\bibnamefont {He}}, \bibinfo {author} {\bibfnamefont {R.}~\bibnamefont {Shaydulin}}, \bibinfo {author} {\bibfnamefont {S.}~\bibnamefont {Chakrabarti}}, \bibinfo {author} {\bibfnamefont {D.}~\bibnamefont {Herman}}, \bibinfo {author} {\bibfnamefont {C.}~\bibnamefont {Li}}, \bibinfo {author} {\bibfnamefont {Y.}~\bibnamefont {Sun}},\ and\ \bibinfo {author} {\bibfnamefont {M.}~\bibnamefont {Pistoia}},\ }\bibfield  {title} {\bibinfo {title} {Alignment between initial state and mixer improves qaoa performance for constrained optimization},\ }\href@noop {} {\bibfield  {journal} {\bibinfo  {journal} {npj Quantum Information}\ }\textbf {\bibinfo {volume} {9}},\ \bibinfo {pages} {121} (\bibinfo {year} {2023})}\BibitemShut {NoStop}%
\bibitem [{\citenamefont {Or{\'u}s}\ \emph {et~al.}(2019)\citenamefont {Or{\'u}s}, \citenamefont {Mugel},\ and\ \citenamefont {Lizaso}}]{orus2019quantum}%
  \BibitemOpen
  \bibfield  {author} {\bibinfo {author} {\bibfnamefont {R.}~\bibnamefont {Or{\'u}s}}, \bibinfo {author} {\bibfnamefont {S.}~\bibnamefont {Mugel}},\ and\ \bibinfo {author} {\bibfnamefont {E.}~\bibnamefont {Lizaso}},\ }\bibfield  {title} {\bibinfo {title} {Quantum computing for finance: Overview and prospects},\ }\href@noop {} {\bibfield  {journal} {\bibinfo  {journal} {Reviews in Physics}\ }\textbf {\bibinfo {volume} {4}},\ \bibinfo {pages} {100028} (\bibinfo {year} {2019})}\BibitemShut {NoStop}%
\bibitem [{\citenamefont {Herman}\ \emph {et~al.}(2022)\citenamefont {Herman}, \citenamefont {Googin}, \citenamefont {Liu}, \citenamefont {Galda}, \citenamefont {Safro}, \citenamefont {Sun}, \citenamefont {Pistoia},\ and\ \citenamefont {Alexeev}}]{herman2022survey}%
  \BibitemOpen
  \bibfield  {author} {\bibinfo {author} {\bibfnamefont {D.}~\bibnamefont {Herman}}, \bibinfo {author} {\bibfnamefont {C.}~\bibnamefont {Googin}}, \bibinfo {author} {\bibfnamefont {X.}~\bibnamefont {Liu}}, \bibinfo {author} {\bibfnamefont {A.}~\bibnamefont {Galda}}, \bibinfo {author} {\bibfnamefont {I.}~\bibnamefont {Safro}}, \bibinfo {author} {\bibfnamefont {Y.}~\bibnamefont {Sun}}, \bibinfo {author} {\bibfnamefont {M.}~\bibnamefont {Pistoia}},\ and\ \bibinfo {author} {\bibfnamefont {Y.}~\bibnamefont {Alexeev}},\ }\bibfield  {title} {\bibinfo {title} {A survey of quantum computing for finance},\ }\href@noop {} {\bibfield  {journal} {\bibinfo  {journal} {arXiv preprint arXiv:2201.02773}\ } (\bibinfo {year} {2022})}\BibitemShut {NoStop}%
\bibitem [{\citenamefont {Biamonte}\ \emph {et~al.}(2017)\citenamefont {Biamonte}, \citenamefont {Wittek}, \citenamefont {Pancotti}, \citenamefont {Rebentrost}, \citenamefont {Wiebe},\ and\ \citenamefont {Lloyd}}]{biamonte2017quantum}%
  \BibitemOpen
  \bibfield  {author} {\bibinfo {author} {\bibfnamefont {J.}~\bibnamefont {Biamonte}}, \bibinfo {author} {\bibfnamefont {P.}~\bibnamefont {Wittek}}, \bibinfo {author} {\bibfnamefont {N.}~\bibnamefont {Pancotti}}, \bibinfo {author} {\bibfnamefont {P.}~\bibnamefont {Rebentrost}}, \bibinfo {author} {\bibfnamefont {N.}~\bibnamefont {Wiebe}},\ and\ \bibinfo {author} {\bibfnamefont {S.}~\bibnamefont {Lloyd}},\ }\bibfield  {title} {\bibinfo {title} {Quantum machine learning},\ }\href@noop {} {\bibfield  {journal} {\bibinfo  {journal} {Nature}\ }\textbf {\bibinfo {volume} {549}},\ \bibinfo {pages} {195} (\bibinfo {year} {2017})}\BibitemShut {NoStop}%
\bibitem [{\citenamefont {Cerezo}\ \emph {et~al.}(2022)\citenamefont {Cerezo}, \citenamefont {Verdon}, \citenamefont {Huang}, \citenamefont {Cincio},\ and\ \citenamefont {Coles}}]{cerezo2022challenges}%
  \BibitemOpen
  \bibfield  {author} {\bibinfo {author} {\bibfnamefont {M.}~\bibnamefont {Cerezo}}, \bibinfo {author} {\bibfnamefont {G.}~\bibnamefont {Verdon}}, \bibinfo {author} {\bibfnamefont {H.-Y.}\ \bibnamefont {Huang}}, \bibinfo {author} {\bibfnamefont {L.}~\bibnamefont {Cincio}},\ and\ \bibinfo {author} {\bibfnamefont {P.~J.}\ \bibnamefont {Coles}},\ }\bibfield  {title} {\bibinfo {title} {Challenges and opportunities in quantum machine learning},\ }\href@noop {} {\bibfield  {journal} {\bibinfo  {journal} {Nature Computational Science}\ }\textbf {\bibinfo {volume} {2}},\ \bibinfo {pages} {567} (\bibinfo {year} {2022})}\BibitemShut {NoStop}%
\bibitem [{\citenamefont {Liu}\ \emph {et~al.}(2023)\citenamefont {Liu}, \citenamefont {Liu}, \citenamefont {Liu}, \citenamefont {Ye}, \citenamefont {Alexeev}, \citenamefont {Eisert},\ and\ \citenamefont {Jiang}}]{liu2023towards}%
  \BibitemOpen
  \bibfield  {author} {\bibinfo {author} {\bibfnamefont {J.}~\bibnamefont {Liu}}, \bibinfo {author} {\bibfnamefont {M.}~\bibnamefont {Liu}}, \bibinfo {author} {\bibfnamefont {J.-P.}\ \bibnamefont {Liu}}, \bibinfo {author} {\bibfnamefont {Z.}~\bibnamefont {Ye}}, \bibinfo {author} {\bibfnamefont {Y.}~\bibnamefont {Alexeev}}, \bibinfo {author} {\bibfnamefont {J.}~\bibnamefont {Eisert}},\ and\ \bibinfo {author} {\bibfnamefont {L.}~\bibnamefont {Jiang}},\ }\bibfield  {title} {\bibinfo {title} {Towards provably efficient quantum algorithms for large-scale machine-learning models},\ }\href@noop {} {\bibfield  {journal} {\bibinfo  {journal} {arXiv preprint arXiv:2303.03428}\ } (\bibinfo {year} {2023})}\BibitemShut {NoStop}%
\bibitem [{\citenamefont {Miessen}\ \emph {et~al.}(2023)\citenamefont {Miessen}, \citenamefont {Ollitrault}, \citenamefont {Tacchino},\ and\ \citenamefont {Tavernelli}}]{miessen2023quantum}%
  \BibitemOpen
  \bibfield  {author} {\bibinfo {author} {\bibfnamefont {A.}~\bibnamefont {Miessen}}, \bibinfo {author} {\bibfnamefont {P.~J.}\ \bibnamefont {Ollitrault}}, \bibinfo {author} {\bibfnamefont {F.}~\bibnamefont {Tacchino}},\ and\ \bibinfo {author} {\bibfnamefont {I.}~\bibnamefont {Tavernelli}},\ }\bibfield  {title} {\bibinfo {title} {Quantum algorithms for quantum dynamics},\ }\href@noop {} {\bibfield  {journal} {\bibinfo  {journal} {Nature Computational Science}\ }\textbf {\bibinfo {volume} {3}},\ \bibinfo {pages} {25} (\bibinfo {year} {2023})}\BibitemShut {NoStop}%
\bibitem [{\citenamefont {Peng}\ \emph {et~al.}(2022)\citenamefont {Peng}, \citenamefont {Gulania}, \citenamefont {Alexeev},\ and\ \citenamefont {Govind}}]{peng2022quantum}%
  \BibitemOpen
  \bibfield  {author} {\bibinfo {author} {\bibfnamefont {B.}~\bibnamefont {Peng}}, \bibinfo {author} {\bibfnamefont {S.}~\bibnamefont {Gulania}}, \bibinfo {author} {\bibfnamefont {Y.}~\bibnamefont {Alexeev}},\ and\ \bibinfo {author} {\bibfnamefont {N.}~\bibnamefont {Govind}},\ }\bibfield  {title} {\bibinfo {title} {Quantum time dynamics employing the {Yang-Baxter} equation for circuit compression},\ }\href@noop {} {\bibfield  {journal} {\bibinfo  {journal} {Physical Review A}\ }\textbf {\bibinfo {volume} {106}},\ \bibinfo {pages} {012412} (\bibinfo {year} {2022})}\BibitemShut {NoStop}%
\bibitem [{\citenamefont {Gulania}\ \emph {et~al.}(2022)\citenamefont {Gulania}, \citenamefont {He}, \citenamefont {Peng}, \citenamefont {Govind},\ and\ \citenamefont {Alexeev}}]{gulania2022quybe}%
  \BibitemOpen
  \bibfield  {author} {\bibinfo {author} {\bibfnamefont {S.}~\bibnamefont {Gulania}}, \bibinfo {author} {\bibfnamefont {Z.}~\bibnamefont {He}}, \bibinfo {author} {\bibfnamefont {B.}~\bibnamefont {Peng}}, \bibinfo {author} {\bibfnamefont {N.}~\bibnamefont {Govind}},\ and\ \bibinfo {author} {\bibfnamefont {Y.}~\bibnamefont {Alexeev}},\ }\bibfield  {title} {\bibinfo {title} {{QuYBE}-an algebraic compiler for quantum circuit compression},\ }in\ \href@noop {} {\emph {\bibinfo {booktitle} {2022 IEEE/ACM 7th Symposium on Edge Computing (SEC)}}}\ (\bibinfo {organization} {IEEE},\ \bibinfo {year} {2022})\ pp.\ \bibinfo {pages} {406--410}\BibitemShut {NoStop}%
\bibitem [{\citenamefont {Fedorov}\ \emph {et~al.}(2022)\citenamefont {Fedorov}, \citenamefont {Peng}, \citenamefont {Govind},\ and\ \citenamefont {Alexeev}}]{fedorov2022vqe}%
  \BibitemOpen
  \bibfield  {author} {\bibinfo {author} {\bibfnamefont {D.~A.}\ \bibnamefont {Fedorov}}, \bibinfo {author} {\bibfnamefont {B.}~\bibnamefont {Peng}}, \bibinfo {author} {\bibfnamefont {N.}~\bibnamefont {Govind}},\ and\ \bibinfo {author} {\bibfnamefont {Y.}~\bibnamefont {Alexeev}},\ }\bibfield  {title} {\bibinfo {title} {{VQE} method: a short survey and recent developments},\ }\href@noop {} {\bibfield  {journal} {\bibinfo  {journal} {Materials Theory}\ }\textbf {\bibinfo {volume} {6}},\ \bibinfo {pages} {1} (\bibinfo {year} {2022})}\BibitemShut {NoStop}%
\bibitem [{\citenamefont {McArdle}\ \emph {et~al.}(2020)\citenamefont {McArdle}, \citenamefont {Endo}, \citenamefont {Aspuru-Guzik}, \citenamefont {Benjamin},\ and\ \citenamefont {Yuan}}]{mcardle2020quantum}%
  \BibitemOpen
  \bibfield  {author} {\bibinfo {author} {\bibfnamefont {S.}~\bibnamefont {McArdle}}, \bibinfo {author} {\bibfnamefont {S.}~\bibnamefont {Endo}}, \bibinfo {author} {\bibfnamefont {A.}~\bibnamefont {Aspuru-Guzik}}, \bibinfo {author} {\bibfnamefont {S.~C.}\ \bibnamefont {Benjamin}},\ and\ \bibinfo {author} {\bibfnamefont {X.}~\bibnamefont {Yuan}},\ }\bibfield  {title} {\bibinfo {title} {Quantum computational chemistry},\ }\href@noop {} {\bibfield  {journal} {\bibinfo  {journal} {Reviews of Modern Physics}\ }\textbf {\bibinfo {volume} {92}},\ \bibinfo {pages} {015003} (\bibinfo {year} {2020})}\BibitemShut {NoStop}%
\bibitem [{\citenamefont {Cao}\ \emph {et~al.}(2019)\citenamefont {Cao}, \citenamefont {Romero}, \citenamefont {Olson}, \citenamefont {Degroote}, \citenamefont {Johnson}, \citenamefont {Kieferov{\'a}}, \citenamefont {Kivlichan}, \citenamefont {Menke}, \citenamefont {Peropadre}, \citenamefont {Sawaya} \emph {et~al.}}]{cao2019quantum}%
  \BibitemOpen
  \bibfield  {author} {\bibinfo {author} {\bibfnamefont {Y.}~\bibnamefont {Cao}}, \bibinfo {author} {\bibfnamefont {J.}~\bibnamefont {Romero}}, \bibinfo {author} {\bibfnamefont {J.~P.}\ \bibnamefont {Olson}}, \bibinfo {author} {\bibfnamefont {M.}~\bibnamefont {Degroote}}, \bibinfo {author} {\bibfnamefont {P.~D.}\ \bibnamefont {Johnson}}, \bibinfo {author} {\bibfnamefont {M.}~\bibnamefont {Kieferov{\'a}}}, \bibinfo {author} {\bibfnamefont {I.~D.}\ \bibnamefont {Kivlichan}}, \bibinfo {author} {\bibfnamefont {T.}~\bibnamefont {Menke}}, \bibinfo {author} {\bibfnamefont {B.}~\bibnamefont {Peropadre}}, \bibinfo {author} {\bibfnamefont {N.~P.}\ \bibnamefont {Sawaya}}, \emph {et~al.},\ }\bibfield  {title} {\bibinfo {title} {Quantum chemistry in the age of quantum computing},\ }\href@noop {} {\bibfield  {journal} {\bibinfo  {journal} {Chemical reviews}\ }\textbf {\bibinfo {volume} {119}},\ \bibinfo {pages} {10856} (\bibinfo {year} {2019})}\BibitemShut {NoStop}%
\bibitem [{\citenamefont {Bittel}\ and\ \citenamefont {Kliesch}(2021)}]{bittel2021training}%
  \BibitemOpen
  \bibfield  {author} {\bibinfo {author} {\bibfnamefont {L.}~\bibnamefont {Bittel}}\ and\ \bibinfo {author} {\bibfnamefont {M.}~\bibnamefont {Kliesch}},\ }\bibfield  {title} {\bibinfo {title} {Training variational quantum algorithms is {NP}-hard},\ }\href@noop {} {\bibfield  {journal} {\bibinfo  {journal} {Physical review letters}\ }\textbf {\bibinfo {volume} {127}},\ \bibinfo {pages} {120502} (\bibinfo {year} {2021})}\BibitemShut {NoStop}%
\bibitem [{\citenamefont {Sung}\ \emph {et~al.}(2020)\citenamefont {Sung}, \citenamefont {Yao}, \citenamefont {Harrigan}, \citenamefont {Rubin}, \citenamefont {Jiang}, \citenamefont {Lin}, \citenamefont {Babbush},\ and\ \citenamefont {McClean}}]{sung2020using}%
  \BibitemOpen
  \bibfield  {author} {\bibinfo {author} {\bibfnamefont {K.~J.}\ \bibnamefont {Sung}}, \bibinfo {author} {\bibfnamefont {J.}~\bibnamefont {Yao}}, \bibinfo {author} {\bibfnamefont {M.~P.}\ \bibnamefont {Harrigan}}, \bibinfo {author} {\bibfnamefont {N.~C.}\ \bibnamefont {Rubin}}, \bibinfo {author} {\bibfnamefont {Z.}~\bibnamefont {Jiang}}, \bibinfo {author} {\bibfnamefont {L.}~\bibnamefont {Lin}}, \bibinfo {author} {\bibfnamefont {R.}~\bibnamefont {Babbush}},\ and\ \bibinfo {author} {\bibfnamefont {J.~R.}\ \bibnamefont {McClean}},\ }\bibfield  {title} {\bibinfo {title} {Using models to improve optimizers for variational quantum algorithms},\ }\href@noop {} {\bibfield  {journal} {\bibinfo  {journal} {Quantum Science and Technology}\ }\textbf {\bibinfo {volume} {5}},\ \bibinfo {pages} {044008} (\bibinfo {year} {2020})}\BibitemShut {NoStop}%
\bibitem [{\citenamefont {Bonet-Monroig}\ \emph {et~al.}(2023)\citenamefont {Bonet-Monroig}, \citenamefont {Wang}, \citenamefont {Vermetten}, \citenamefont {Senjean}, \citenamefont {Moussa}, \citenamefont {B{\"a}ck}, \citenamefont {Dunjko},\ and\ \citenamefont {O'Brien}}]{bonet2023performance}%
  \BibitemOpen
  \bibfield  {author} {\bibinfo {author} {\bibfnamefont {X.}~\bibnamefont {Bonet-Monroig}}, \bibinfo {author} {\bibfnamefont {H.}~\bibnamefont {Wang}}, \bibinfo {author} {\bibfnamefont {D.}~\bibnamefont {Vermetten}}, \bibinfo {author} {\bibfnamefont {B.}~\bibnamefont {Senjean}}, \bibinfo {author} {\bibfnamefont {C.}~\bibnamefont {Moussa}}, \bibinfo {author} {\bibfnamefont {T.}~\bibnamefont {B{\"a}ck}}, \bibinfo {author} {\bibfnamefont {V.}~\bibnamefont {Dunjko}},\ and\ \bibinfo {author} {\bibfnamefont {T.~E.}\ \bibnamefont {O'Brien}},\ }\bibfield  {title} {\bibinfo {title} {Performance comparison of optimization methods on variational quantum algorithms},\ }\href@noop {} {\bibfield  {journal} {\bibinfo  {journal} {Physical Review A}\ }\textbf {\bibinfo {volume} {107}},\ \bibinfo {pages} {032407} (\bibinfo {year} {2023})}\BibitemShut {NoStop}%
\bibitem [{\citenamefont {Gily{\'e}n}\ \emph {et~al.}(2019)\citenamefont {Gily{\'e}n}, \citenamefont {Arunachalam},\ and\ \citenamefont {Wiebe}}]{gilyen2019optimizing}%
  \BibitemOpen
  \bibfield  {author} {\bibinfo {author} {\bibfnamefont {A.}~\bibnamefont {Gily{\'e}n}}, \bibinfo {author} {\bibfnamefont {S.}~\bibnamefont {Arunachalam}},\ and\ \bibinfo {author} {\bibfnamefont {N.}~\bibnamefont {Wiebe}},\ }\bibfield  {title} {\bibinfo {title} {Optimizing quantum optimization algorithms via faster quantum gradient computation},\ }in\ \href@noop {} {\emph {\bibinfo {booktitle} {Proceedings of the Thirtieth Annual ACM-SIAM Symposium on Discrete Algorithms}}}\ (\bibinfo {organization} {SIAM},\ \bibinfo {year} {2019})\ pp.\ \bibinfo {pages} {1425--1444}\BibitemShut {NoStop}%
\bibitem [{\citenamefont {Shaydulin}\ \emph {et~al.}(2023)\citenamefont {Shaydulin}, \citenamefont {Lotshaw}, \citenamefont {Larson}, \citenamefont {Ostrowski},\ and\ \citenamefont {Humble}}]{shaydulin2023parameter}%
  \BibitemOpen
  \bibfield  {author} {\bibinfo {author} {\bibfnamefont {R.}~\bibnamefont {Shaydulin}}, \bibinfo {author} {\bibfnamefont {P.~C.}\ \bibnamefont {Lotshaw}}, \bibinfo {author} {\bibfnamefont {J.}~\bibnamefont {Larson}}, \bibinfo {author} {\bibfnamefont {J.}~\bibnamefont {Ostrowski}},\ and\ \bibinfo {author} {\bibfnamefont {T.~S.}\ \bibnamefont {Humble}},\ }\bibfield  {title} {\bibinfo {title} {Parameter transfer for quantum approximate optimization of weighted maxcut},\ }\href@noop {} {\bibfield  {journal} {\bibinfo  {journal} {ACM Transactions on Quantum Computing}\ }\textbf {\bibinfo {volume} {4}},\ \bibinfo {pages} {1} (\bibinfo {year} {2023})}\BibitemShut {NoStop}%
\bibitem [{\citenamefont {Stilck~Fran{\c{c}}a}\ and\ \citenamefont {Garcia-Patron}(2021)}]{stilck2021limitations}%
  \BibitemOpen
  \bibfield  {author} {\bibinfo {author} {\bibfnamefont {D.}~\bibnamefont {Stilck~Fran{\c{c}}a}}\ and\ \bibinfo {author} {\bibfnamefont {R.}~\bibnamefont {Garcia-Patron}},\ }\bibfield  {title} {\bibinfo {title} {Limitations of optimization algorithms on noisy quantum devices},\ }\href@noop {} {\bibfield  {journal} {\bibinfo  {journal} {Nature Physics}\ }\textbf {\bibinfo {volume} {17}},\ \bibinfo {pages} {1221} (\bibinfo {year} {2021})}\BibitemShut {NoStop}%
\bibitem [{\citenamefont {Gonz{\'a}lez-Garc{\'\i}a}\ \emph {et~al.}(2022)\citenamefont {Gonz{\'a}lez-Garc{\'\i}a}, \citenamefont {Trivedi},\ and\ \citenamefont {Cirac}}]{gonzalez2022error}%
  \BibitemOpen
  \bibfield  {author} {\bibinfo {author} {\bibfnamefont {G.}~\bibnamefont {Gonz{\'a}lez-Garc{\'\i}a}}, \bibinfo {author} {\bibfnamefont {R.}~\bibnamefont {Trivedi}},\ and\ \bibinfo {author} {\bibfnamefont {J.~I.}\ \bibnamefont {Cirac}},\ }\bibfield  {title} {\bibinfo {title} {Error propagation in {NISQ} devices for solving classical optimization problems},\ }\href@noop {} {\bibfield  {journal} {\bibinfo  {journal} {PRX Quantum}\ }\textbf {\bibinfo {volume} {3}},\ \bibinfo {pages} {040326} (\bibinfo {year} {2022})}\BibitemShut {NoStop}%
\bibitem [{\citenamefont {De~Palma}\ \emph {et~al.}(2023)\citenamefont {De~Palma}, \citenamefont {Marvian}, \citenamefont {Rouz{\'e}},\ and\ \citenamefont {Fran{\c{c}}a}}]{de2023limitations}%
  \BibitemOpen
  \bibfield  {author} {\bibinfo {author} {\bibfnamefont {G.}~\bibnamefont {De~Palma}}, \bibinfo {author} {\bibfnamefont {M.}~\bibnamefont {Marvian}}, \bibinfo {author} {\bibfnamefont {C.}~\bibnamefont {Rouz{\'e}}},\ and\ \bibinfo {author} {\bibfnamefont {D.~S.}\ \bibnamefont {Fran{\c{c}}a}},\ }\bibfield  {title} {\bibinfo {title} {Limitations of variational quantum algorithms: a quantum optimal transport approach},\ }\href@noop {} {\bibfield  {journal} {\bibinfo  {journal} {PRX Quantum}\ }\textbf {\bibinfo {volume} {4}},\ \bibinfo {pages} {010309} (\bibinfo {year} {2023})}\BibitemShut {NoStop}%
\bibitem [{\citenamefont {Harper}\ \emph {et~al.}(2020)\citenamefont {Harper}, \citenamefont {Flammia},\ and\ \citenamefont {Wallman}}]{harper2020efficient}%
  \BibitemOpen
  \bibfield  {author} {\bibinfo {author} {\bibfnamefont {R.}~\bibnamefont {Harper}}, \bibinfo {author} {\bibfnamefont {S.~T.}\ \bibnamefont {Flammia}},\ and\ \bibinfo {author} {\bibfnamefont {J.~J.}\ \bibnamefont {Wallman}},\ }\bibfield  {title} {\bibinfo {title} {Efficient learning of quantum noise},\ }\href@noop {} {\bibfield  {journal} {\bibinfo  {journal} {Nature Physics}\ }\textbf {\bibinfo {volume} {16}},\ \bibinfo {pages} {1184} (\bibinfo {year} {2020})}\BibitemShut {NoStop}%
\bibitem [{\citenamefont {Endo}\ \emph {et~al.}(2018)\citenamefont {Endo}, \citenamefont {Benjamin},\ and\ \citenamefont {Li}}]{endo2018practical}%
  \BibitemOpen
  \bibfield  {author} {\bibinfo {author} {\bibfnamefont {S.}~\bibnamefont {Endo}}, \bibinfo {author} {\bibfnamefont {S.~C.}\ \bibnamefont {Benjamin}},\ and\ \bibinfo {author} {\bibfnamefont {Y.}~\bibnamefont {Li}},\ }\bibfield  {title} {\bibinfo {title} {Practical quantum error mitigation for near-future applications},\ }\href@noop {} {\bibfield  {journal} {\bibinfo  {journal} {Physical Review X}\ }\textbf {\bibinfo {volume} {8}},\ \bibinfo {pages} {031027} (\bibinfo {year} {2018})}\BibitemShut {NoStop}%
\bibitem [{\citenamefont {Suzuki}\ \emph {et~al.}(2022)\citenamefont {Suzuki}, \citenamefont {Endo}, \citenamefont {Fujii},\ and\ \citenamefont {Tokunaga}}]{suzuki2022quantum}%
  \BibitemOpen
  \bibfield  {author} {\bibinfo {author} {\bibfnamefont {Y.}~\bibnamefont {Suzuki}}, \bibinfo {author} {\bibfnamefont {S.}~\bibnamefont {Endo}}, \bibinfo {author} {\bibfnamefont {K.}~\bibnamefont {Fujii}},\ and\ \bibinfo {author} {\bibfnamefont {Y.}~\bibnamefont {Tokunaga}},\ }\bibfield  {title} {\bibinfo {title} {Quantum error mitigation as a universal error reduction technique: applications from the {NISQ} to the fault-tolerant quantum computing eras},\ }\href@noop {} {\bibfield  {journal} {\bibinfo  {journal} {PRX Quantum}\ }\textbf {\bibinfo {volume} {3}},\ \bibinfo {pages} {010345} (\bibinfo {year} {2022})}\BibitemShut {NoStop}%
\bibitem [{\citenamefont {Liu}\ and\ \citenamefont {Zhou}(2020)}]{liu2020reliability}%
  \BibitemOpen
  \bibfield  {author} {\bibinfo {author} {\bibfnamefont {J.}~\bibnamefont {Liu}}\ and\ \bibinfo {author} {\bibfnamefont {H.}~\bibnamefont {Zhou}},\ }\bibfield  {title} {\bibinfo {title} {Reliability modeling of {NISQ}-era quantum computers},\ }in\ \href@noop {} {\emph {\bibinfo {booktitle} {2020 IEEE international symposium on workload characterization (IISWC)}}}\ (\bibinfo {organization} {IEEE},\ \bibinfo {year} {2020})\ pp.\ \bibinfo {pages} {94--105}\BibitemShut {NoStop}%
\bibitem [{\citenamefont {Wang}\ \emph {et~al.}(2022)\citenamefont {Wang}, \citenamefont {Ding}, \citenamefont {Gu}, \citenamefont {Lin}, \citenamefont {Pan}, \citenamefont {Chong},\ and\ \citenamefont {Han}}]{wang2022quantumnas}%
  \BibitemOpen
  \bibfield  {author} {\bibinfo {author} {\bibfnamefont {H.}~\bibnamefont {Wang}}, \bibinfo {author} {\bibfnamefont {Y.}~\bibnamefont {Ding}}, \bibinfo {author} {\bibfnamefont {J.}~\bibnamefont {Gu}}, \bibinfo {author} {\bibfnamefont {Y.}~\bibnamefont {Lin}}, \bibinfo {author} {\bibfnamefont {D.~Z.}\ \bibnamefont {Pan}}, \bibinfo {author} {\bibfnamefont {F.~T.}\ \bibnamefont {Chong}},\ and\ \bibinfo {author} {\bibfnamefont {S.}~\bibnamefont {Han}},\ }\bibfield  {title} {\bibinfo {title} {{QuantumNAS}: Noise-adaptive search for robust quantum circuits},\ }in\ \href@noop {} {\emph {\bibinfo {booktitle} {2022 IEEE International Symposium on High-Performance Computer Architecture (HPCA)}}}\ (\bibinfo {organization} {IEEE},\ \bibinfo {year} {2022})\ pp.\ \bibinfo {pages} {692--708}\BibitemShut {NoStop}%
\bibitem [{\citenamefont {Magann}\ \emph {et~al.}(2021)\citenamefont {Magann}, \citenamefont {Arenz}, \citenamefont {Grace}, \citenamefont {Ho}, \citenamefont {Kosut}, \citenamefont {McClean}, \citenamefont {Rabitz},\ and\ \citenamefont {Sarovar}}]{magann2021pulses}%
  \BibitemOpen
  \bibfield  {author} {\bibinfo {author} {\bibfnamefont {A.~B.}\ \bibnamefont {Magann}}, \bibinfo {author} {\bibfnamefont {C.}~\bibnamefont {Arenz}}, \bibinfo {author} {\bibfnamefont {M.~D.}\ \bibnamefont {Grace}}, \bibinfo {author} {\bibfnamefont {T.-S.}\ \bibnamefont {Ho}}, \bibinfo {author} {\bibfnamefont {R.~L.}\ \bibnamefont {Kosut}}, \bibinfo {author} {\bibfnamefont {J.~R.}\ \bibnamefont {McClean}}, \bibinfo {author} {\bibfnamefont {H.~A.}\ \bibnamefont {Rabitz}},\ and\ \bibinfo {author} {\bibfnamefont {M.}~\bibnamefont {Sarovar}},\ }\bibfield  {title} {\bibinfo {title} {From pulses to circuits and back again: A quantum optimal control perspective on variational quantum algorithms},\ }\href@noop {} {\bibfield  {journal} {\bibinfo  {journal} {PRX Quantum}\ }\textbf {\bibinfo {volume} {2}},\ \bibinfo {pages} {010101} (\bibinfo {year} {2021})}\BibitemShut {NoStop}%
\bibitem [{\citenamefont {Liang}\ \emph {et~al.}(2022)\citenamefont {Liang}, \citenamefont {Song}, \citenamefont {Cheng}, \citenamefont {He}, \citenamefont {Liu}, \citenamefont {Wang}, \citenamefont {Qin}, \citenamefont {Wang}, \citenamefont {Han}, \citenamefont {Qian} \emph {et~al.}}]{liang2022hybrid}%
  \BibitemOpen
  \bibfield  {author} {\bibinfo {author} {\bibfnamefont {Z.}~\bibnamefont {Liang}}, \bibinfo {author} {\bibfnamefont {Z.}~\bibnamefont {Song}}, \bibinfo {author} {\bibfnamefont {J.}~\bibnamefont {Cheng}}, \bibinfo {author} {\bibfnamefont {Z.}~\bibnamefont {He}}, \bibinfo {author} {\bibfnamefont {J.}~\bibnamefont {Liu}}, \bibinfo {author} {\bibfnamefont {H.}~\bibnamefont {Wang}}, \bibinfo {author} {\bibfnamefont {R.}~\bibnamefont {Qin}}, \bibinfo {author} {\bibfnamefont {Y.}~\bibnamefont {Wang}}, \bibinfo {author} {\bibfnamefont {S.}~\bibnamefont {Han}}, \bibinfo {author} {\bibfnamefont {X.}~\bibnamefont {Qian}}, \emph {et~al.},\ }\bibfield  {title} {\bibinfo {title} {Hybrid gate-pulse model for variational quantum algorithms},\ }\href@noop {} {\bibfield  {journal} {\bibinfo  {journal} {arXiv preprint arXiv:2212.00661}\ } (\bibinfo {year} {2022})}\BibitemShut {NoStop}%
\bibitem [{\citenamefont {Burnett}\ \emph {et~al.}(2019)\citenamefont {Burnett}, \citenamefont {Bengtsson}, \citenamefont {Scigliuzzo}, \citenamefont {Niepce}, \citenamefont {Kudra}, \citenamefont {Delsing},\ and\ \citenamefont {Bylander}}]{burnett2019decoherence}%
  \BibitemOpen
  \bibfield  {author} {\bibinfo {author} {\bibfnamefont {J.~J.}\ \bibnamefont {Burnett}}, \bibinfo {author} {\bibfnamefont {A.}~\bibnamefont {Bengtsson}}, \bibinfo {author} {\bibfnamefont {M.}~\bibnamefont {Scigliuzzo}}, \bibinfo {author} {\bibfnamefont {D.}~\bibnamefont {Niepce}}, \bibinfo {author} {\bibfnamefont {M.}~\bibnamefont {Kudra}}, \bibinfo {author} {\bibfnamefont {P.}~\bibnamefont {Delsing}},\ and\ \bibinfo {author} {\bibfnamefont {J.}~\bibnamefont {Bylander}},\ }\bibfield  {title} {\bibinfo {title} {Decoherence benchmarking of superconducting qubits},\ }\href@noop {} {\bibfield  {journal} {\bibinfo  {journal} {npj Quantum Information}\ }\textbf {\bibinfo {volume} {5}},\ \bibinfo {pages} {54} (\bibinfo {year} {2019})}\BibitemShut {NoStop}%
\bibitem [{\citenamefont {Dasgupta}\ and\ \citenamefont {Humble}(2022{\natexlab{a}})}]{dasgupta2022assessing}%
  \BibitemOpen
  \bibfield  {author} {\bibinfo {author} {\bibfnamefont {S.}~\bibnamefont {Dasgupta}}\ and\ \bibinfo {author} {\bibfnamefont {T.~S.}\ \bibnamefont {Humble}},\ }\bibfield  {title} {\bibinfo {title} {Assessing the stability of noisy quantum computation},\ }in\ \href@noop {} {\emph {\bibinfo {booktitle} {Quantum Communications and Quantum Imaging XX}}},\ Vol.\ \bibinfo {volume} {12238}\ (\bibinfo {organization} {SPIE},\ \bibinfo {year} {2022})\ pp.\ \bibinfo {pages} {44--49}\BibitemShut {NoStop}%
\bibitem [{\citenamefont {Dasgupta}\ and\ \citenamefont {Humble}(2022{\natexlab{b}})}]{dasgupta2022characterizing}%
  \BibitemOpen
  \bibfield  {author} {\bibinfo {author} {\bibfnamefont {S.}~\bibnamefont {Dasgupta}}\ and\ \bibinfo {author} {\bibfnamefont {T.~S.}\ \bibnamefont {Humble}},\ }\bibfield  {title} {\bibinfo {title} {Characterizing the reproducibility of noisy quantum circuits},\ }\href@noop {} {\bibfield  {journal} {\bibinfo  {journal} {Entropy}\ }\textbf {\bibinfo {volume} {24}},\ \bibinfo {pages} {244} (\bibinfo {year} {2022}{\natexlab{b}})}\BibitemShut {NoStop}%
\bibitem [{\citenamefont {Hu}\ \emph {et~al.}(2023)\citenamefont {Hu}, \citenamefont {Wolle}, \citenamefont {Tian}, \citenamefont {Guan}, \citenamefont {Humble},\ and\ \citenamefont {Jiang}}]{hu2023toward}%
  \BibitemOpen
  \bibfield  {author} {\bibinfo {author} {\bibfnamefont {Z.}~\bibnamefont {Hu}}, \bibinfo {author} {\bibfnamefont {R.}~\bibnamefont {Wolle}}, \bibinfo {author} {\bibfnamefont {M.}~\bibnamefont {Tian}}, \bibinfo {author} {\bibfnamefont {Q.}~\bibnamefont {Guan}}, \bibinfo {author} {\bibfnamefont {T.}~\bibnamefont {Humble}},\ and\ \bibinfo {author} {\bibfnamefont {W.}~\bibnamefont {Jiang}},\ }\bibfield  {title} {\bibinfo {title} {Toward consistent high-fidelity quantum learning on unstable devices via efficient in-situ calibration},\ }in\ \href@noop {} {\emph {\bibinfo {booktitle} {2023 IEEE International Conference on Quantum Computing and Engineering (QCE)}}},\ Vol.~\bibinfo {volume} {1}\ (\bibinfo {organization} {IEEE},\ \bibinfo {year} {2023})\ pp.\ \bibinfo {pages} {848--858}\BibitemShut {NoStop}%
\bibitem [{\citenamefont {Zhang}\ \emph {et~al.}(2023)\citenamefont {Zhang}, \citenamefont {Wang}, \citenamefont {Ravi}, \citenamefont {Chong}, \citenamefont {Han}, \citenamefont {Mueller},\ and\ \citenamefont {Chen}}]{zhang2023disq}%
  \BibitemOpen
  \bibfield  {author} {\bibinfo {author} {\bibfnamefont {J.}~\bibnamefont {Zhang}}, \bibinfo {author} {\bibfnamefont {H.}~\bibnamefont {Wang}}, \bibinfo {author} {\bibfnamefont {G.~S.}\ \bibnamefont {Ravi}}, \bibinfo {author} {\bibfnamefont {F.~T.}\ \bibnamefont {Chong}}, \bibinfo {author} {\bibfnamefont {S.}~\bibnamefont {Han}}, \bibinfo {author} {\bibfnamefont {F.}~\bibnamefont {Mueller}},\ and\ \bibinfo {author} {\bibfnamefont {Y.}~\bibnamefont {Chen}},\ }\bibfield  {title} {\bibinfo {title} {{Disq}: Dynamic iteration skipping for variational quantum algorithms},\ }in\ \href@noop {} {\emph {\bibinfo {booktitle} {2023 IEEE International Conference on Quantum Computing and Engineering (QCE)}}},\ Vol.~\bibinfo {volume} {1}\ (\bibinfo {organization} {IEEE},\ \bibinfo {year} {2023})\ pp.\ \bibinfo {pages} {1062--1073}\BibitemShut {NoStop}%
\bibitem [{\citenamefont {Dasgupta}\ and\ \citenamefont {Humble}(2023)}]{dasgupta2023reliable}%
  \BibitemOpen
  \bibfield  {author} {\bibinfo {author} {\bibfnamefont {S.}~\bibnamefont {Dasgupta}}\ and\ \bibinfo {author} {\bibfnamefont {T.~S.}\ \bibnamefont {Humble}},\ }\bibfield  {title} {\bibinfo {title} {Reliable devices yield stable quantum computations},\ }\href@noop {} {\bibfield  {journal} {\bibinfo  {journal} {arXiv preprint arXiv:2307.05381}\ } (\bibinfo {year} {2023})}\BibitemShut {NoStop}%
\bibitem [{\citenamefont {Dasgupta}\ \emph {et~al.}(2023)\citenamefont {Dasgupta}, \citenamefont {Humble},\ and\ \citenamefont {Danageozian}}]{dasgupta2023adaptive}%
  \BibitemOpen
  \bibfield  {author} {\bibinfo {author} {\bibfnamefont {S.}~\bibnamefont {Dasgupta}}, \bibinfo {author} {\bibfnamefont {T.~S.}\ \bibnamefont {Humble}},\ and\ \bibinfo {author} {\bibfnamefont {A.}~\bibnamefont {Danageozian}},\ }\bibfield  {title} {\bibinfo {title} {Adaptive mitigation of time-varying quantum noise},\ }in\ \href@noop {} {\emph {\bibinfo {booktitle} {2023 IEEE International Conference on Quantum Computing and Engineering (QCE)}}},\ Vol.~\bibinfo {volume} {1}\ (\bibinfo {organization} {IEEE},\ \bibinfo {year} {2023})\ pp.\ \bibinfo {pages} {99--110}\BibitemShut {NoStop}%
\bibitem [{\citenamefont {Ravi}\ \emph {et~al.}(2023)\citenamefont {Ravi}, \citenamefont {Smith}, \citenamefont {Baker}, \citenamefont {Kannan}, \citenamefont {Earnest}, \citenamefont {Javadi-Abhari}, \citenamefont {Hoffmann},\ and\ \citenamefont {Chong}}]{ravi2023navigating}%
  \BibitemOpen
  \bibfield  {author} {\bibinfo {author} {\bibfnamefont {G.~S.}\ \bibnamefont {Ravi}}, \bibinfo {author} {\bibfnamefont {K.}~\bibnamefont {Smith}}, \bibinfo {author} {\bibfnamefont {J.~M.}\ \bibnamefont {Baker}}, \bibinfo {author} {\bibfnamefont {T.}~\bibnamefont {Kannan}}, \bibinfo {author} {\bibfnamefont {N.}~\bibnamefont {Earnest}}, \bibinfo {author} {\bibfnamefont {A.}~\bibnamefont {Javadi-Abhari}}, \bibinfo {author} {\bibfnamefont {H.}~\bibnamefont {Hoffmann}},\ and\ \bibinfo {author} {\bibfnamefont {F.~T.}\ \bibnamefont {Chong}},\ }\bibfield  {title} {\bibinfo {title} {Navigating the dynamic noise landscape of variational quantum algorithms with {QISMET}},\ }in\ \href@noop {} {\emph {\bibinfo {booktitle} {Proceedings of the 28th ACM International Conference on Architectural Support for Programming Languages and Operating Systems, Volume 2}}}\ (\bibinfo {year} {2023})\ pp.\ \bibinfo {pages} {515--529}\BibitemShut {NoStop}%
\bibitem [{\citenamefont {Rahimian}\ and\ \citenamefont {Mehrotra}(2019)}]{rahimian2019distributionally}%
  \BibitemOpen
  \bibfield  {author} {\bibinfo {author} {\bibfnamefont {H.}~\bibnamefont {Rahimian}}\ and\ \bibinfo {author} {\bibfnamefont {S.}~\bibnamefont {Mehrotra}},\ }\bibfield  {title} {\bibinfo {title} {Distributionally robust optimization: A review},\ }\href@noop {} {\bibfield  {journal} {\bibinfo  {journal} {arXiv preprint arXiv:1908.05659}\ } (\bibinfo {year} {2019})}\BibitemShut {NoStop}%
\bibitem [{\citenamefont {Lin}\ \emph {et~al.}(2022)\citenamefont {Lin}, \citenamefont {Fang},\ and\ \citenamefont {Gao}}]{lin2022distributionally}%
  \BibitemOpen
  \bibfield  {author} {\bibinfo {author} {\bibfnamefont {F.}~\bibnamefont {Lin}}, \bibinfo {author} {\bibfnamefont {X.}~\bibnamefont {Fang}},\ and\ \bibinfo {author} {\bibfnamefont {Z.}~\bibnamefont {Gao}},\ }\bibfield  {title} {\bibinfo {title} {Distributionally robust optimization: A review on theory and applications},\ }\href@noop {} {\bibfield  {journal} {\bibinfo  {journal} {Numerical Algebra, Control and Optimization}\ }\textbf {\bibinfo {volume} {12}},\ \bibinfo {pages} {159} (\bibinfo {year} {2022})}\BibitemShut {NoStop}%
\bibitem [{\citenamefont {Kuhn}\ \emph {et~al.}(2019)\citenamefont {Kuhn}, \citenamefont {Esfahani}, \citenamefont {Nguyen},\ and\ \citenamefont {Shafieezadeh-Abadeh}}]{kuhn2019wasserstein}%
  \BibitemOpen
  \bibfield  {author} {\bibinfo {author} {\bibfnamefont {D.}~\bibnamefont {Kuhn}}, \bibinfo {author} {\bibfnamefont {P.~M.}\ \bibnamefont {Esfahani}}, \bibinfo {author} {\bibfnamefont {V.~A.}\ \bibnamefont {Nguyen}},\ and\ \bibinfo {author} {\bibfnamefont {S.}~\bibnamefont {Shafieezadeh-Abadeh}},\ }\bibfield  {title} {\bibinfo {title} {Wasserstein distributionally robust optimization: Theory and applications in machine learning},\ }in\ \href@noop {} {\emph {\bibinfo {booktitle} {Operations research \& management science in the age of analytics}}}\ (\bibinfo  {publisher} {Informs},\ \bibinfo {year} {2019})\ pp.\ \bibinfo {pages} {130--166}\BibitemShut {NoStop}%
\bibitem [{\citenamefont {Delage}\ and\ \citenamefont {Ye}(2010)}]{delage2010distributionally}%
  \BibitemOpen
  \bibfield  {author} {\bibinfo {author} {\bibfnamefont {E.}~\bibnamefont {Delage}}\ and\ \bibinfo {author} {\bibfnamefont {Y.}~\bibnamefont {Ye}},\ }\bibfield  {title} {\bibinfo {title} {Distributionally robust optimization under moment uncertainty with application to data-driven problems},\ }\href@noop {} {\bibfield  {journal} {\bibinfo  {journal} {Operations research}\ }\textbf {\bibinfo {volume} {58}},\ \bibinfo {pages} {595} (\bibinfo {year} {2010})}\BibitemShut {NoStop}%
\bibitem [{\citenamefont {Kirschner}\ \emph {et~al.}(2020)\citenamefont {Kirschner}, \citenamefont {Bogunovic}, \citenamefont {Jegelka},\ and\ \citenamefont {Krause}}]{kirschner2020distributionally}%
  \BibitemOpen
  \bibfield  {author} {\bibinfo {author} {\bibfnamefont {J.}~\bibnamefont {Kirschner}}, \bibinfo {author} {\bibfnamefont {I.}~\bibnamefont {Bogunovic}}, \bibinfo {author} {\bibfnamefont {S.}~\bibnamefont {Jegelka}},\ and\ \bibinfo {author} {\bibfnamefont {A.}~\bibnamefont {Krause}},\ }\bibfield  {title} {\bibinfo {title} {Distributionally robust {Bayesian} optimization},\ }in\ \href@noop {} {\emph {\bibinfo {booktitle} {International Conference on Artificial Intelligence and Statistics}}}\ (\bibinfo {organization} {PMLR},\ \bibinfo {year} {2020})\ pp.\ \bibinfo {pages} {2174--2184}\BibitemShut {NoStop}%
\bibitem [{\citenamefont {Tay}\ \emph {et~al.}(2022)\citenamefont {Tay}, \citenamefont {Foo}, \citenamefont {Daisuke}, \citenamefont {Leong},\ and\ \citenamefont {Low}}]{tay2022efficient}%
  \BibitemOpen
  \bibfield  {author} {\bibinfo {author} {\bibfnamefont {S.~S.}\ \bibnamefont {Tay}}, \bibinfo {author} {\bibfnamefont {C.~S.}\ \bibnamefont {Foo}}, \bibinfo {author} {\bibfnamefont {U.}~\bibnamefont {Daisuke}}, \bibinfo {author} {\bibfnamefont {R.}~\bibnamefont {Leong}},\ and\ \bibinfo {author} {\bibfnamefont {B.~K.~H.}\ \bibnamefont {Low}},\ }\bibfield  {title} {\bibinfo {title} {Efficient distributionally robust {Bayesian} optimization with worst-case sensitivity},\ }in\ \href@noop {} {\emph {\bibinfo {booktitle} {International Conference on Machine Learning}}}\ (\bibinfo {organization} {PMLR},\ \bibinfo {year} {2022})\ pp.\ \bibinfo {pages} {21180--21204}\BibitemShut {NoStop}%
\bibitem [{\citenamefont {Husain}\ \emph {et~al.}(2022)\citenamefont {Husain}, \citenamefont {Nguyen},\ and\ \citenamefont {Hengel}}]{husain2022distributionally}%
  \BibitemOpen
  \bibfield  {author} {\bibinfo {author} {\bibfnamefont {H.}~\bibnamefont {Husain}}, \bibinfo {author} {\bibfnamefont {V.}~\bibnamefont {Nguyen}},\ and\ \bibinfo {author} {\bibfnamefont {A.~v.~d.}\ \bibnamefont {Hengel}},\ }\bibfield  {title} {\bibinfo {title} {Distributionally robust {Bayesian} optimization with $\phi$-divergences},\ }\href@noop {} {\bibfield  {journal} {\bibinfo  {journal} {arXiv preprint arXiv:2203.02128}\ } (\bibinfo {year} {2022})}\BibitemShut {NoStop}%
\bibitem [{\citenamefont {Iannelli}\ and\ \citenamefont {Jansen}(2021)}]{iannelli2021noisy}%
  \BibitemOpen
  \bibfield  {author} {\bibinfo {author} {\bibfnamefont {G.}~\bibnamefont {Iannelli}}\ and\ \bibinfo {author} {\bibfnamefont {K.}~\bibnamefont {Jansen}},\ }\bibfield  {title} {\bibinfo {title} {Noisy {Bayesian} optimization for variational quantum eigensolvers},\ }\href@noop {} {\bibfield  {journal} {\bibinfo  {journal} {arXiv preprint arXiv:2112.00426}\ } (\bibinfo {year} {2021})}\BibitemShut {NoStop}%
\bibitem [{\citenamefont {Duffield}\ \emph {et~al.}(2023)\citenamefont {Duffield}, \citenamefont {Benedetti},\ and\ \citenamefont {Rosenkranz}}]{duffield2023bayesian}%
  \BibitemOpen
  \bibfield  {author} {\bibinfo {author} {\bibfnamefont {S.}~\bibnamefont {Duffield}}, \bibinfo {author} {\bibfnamefont {M.}~\bibnamefont {Benedetti}},\ and\ \bibinfo {author} {\bibfnamefont {M.}~\bibnamefont {Rosenkranz}},\ }\bibfield  {title} {\bibinfo {title} {Bayesian learning of parameterised quantum circuits},\ }\href@noop {} {\bibfield  {journal} {\bibinfo  {journal} {Machine Learning: Science and Technology}\ }\textbf {\bibinfo {volume} {4}},\ \bibinfo {pages} {025007} (\bibinfo {year} {2023})}\BibitemShut {NoStop}%
\bibitem [{\citenamefont {Self}\ \emph {et~al.}(2021)\citenamefont {Self}, \citenamefont {Khosla}, \citenamefont {Smith}, \citenamefont {Sauvage}, \citenamefont {Haynes}, \citenamefont {Knolle}, \citenamefont {Mintert},\ and\ \citenamefont {Kim}}]{self2021variational}%
  \BibitemOpen
  \bibfield  {author} {\bibinfo {author} {\bibfnamefont {C.~N.}\ \bibnamefont {Self}}, \bibinfo {author} {\bibfnamefont {K.~E.}\ \bibnamefont {Khosla}}, \bibinfo {author} {\bibfnamefont {A.~W.}\ \bibnamefont {Smith}}, \bibinfo {author} {\bibfnamefont {F.}~\bibnamefont {Sauvage}}, \bibinfo {author} {\bibfnamefont {P.~D.}\ \bibnamefont {Haynes}}, \bibinfo {author} {\bibfnamefont {J.}~\bibnamefont {Knolle}}, \bibinfo {author} {\bibfnamefont {F.}~\bibnamefont {Mintert}},\ and\ \bibinfo {author} {\bibfnamefont {M.}~\bibnamefont {Kim}},\ }\bibfield  {title} {\bibinfo {title} {Variational quantum algorithm with information sharing},\ }\href@noop {} {\bibfield  {journal} {\bibinfo  {journal} {npj Quantum Information}\ }\textbf {\bibinfo {volume} {7}},\ \bibinfo {pages} {1} (\bibinfo {year} {2021})}\BibitemShut {NoStop}%
\bibitem [{\citenamefont {M{\"u}ller}\ \emph {et~al.}(2022)\citenamefont {M{\"u}ller}, \citenamefont {Lavrijsen}, \citenamefont {Iancu},\ and\ \citenamefont {de~Jong}}]{muller2022accelerating}%
  \BibitemOpen
  \bibfield  {author} {\bibinfo {author} {\bibfnamefont {J.}~\bibnamefont {M{\"u}ller}}, \bibinfo {author} {\bibfnamefont {W.}~\bibnamefont {Lavrijsen}}, \bibinfo {author} {\bibfnamefont {C.}~\bibnamefont {Iancu}},\ and\ \bibinfo {author} {\bibfnamefont {W.}~\bibnamefont {de~Jong}},\ }\bibfield  {title} {\bibinfo {title} {Accelerating noisy {VQE} optimization with {Gaussian} processes},\ }in\ \href@noop {} {\emph {\bibinfo {booktitle} {2022 IEEE International Conference on Quantum Computing and Engineering (QCE)}}}\ (\bibinfo {organization} {IEEE},\ \bibinfo {year} {2022})\ pp.\ \bibinfo {pages} {215--225}\BibitemShut {NoStop}%
\bibitem [{\citenamefont {Tibaldi}\ \emph {et~al.}(2023)\citenamefont {Tibaldi}, \citenamefont {Vodola}, \citenamefont {Tignone},\ and\ \citenamefont {Ercolessi}}]{tibaldi2023bayesian}%
  \BibitemOpen
  \bibfield  {author} {\bibinfo {author} {\bibfnamefont {S.}~\bibnamefont {Tibaldi}}, \bibinfo {author} {\bibfnamefont {D.}~\bibnamefont {Vodola}}, \bibinfo {author} {\bibfnamefont {E.}~\bibnamefont {Tignone}},\ and\ \bibinfo {author} {\bibfnamefont {E.}~\bibnamefont {Ercolessi}},\ }\bibfield  {title} {\bibinfo {title} {Bayesian optimization for {QAOA}},\ }\href@noop {} {\bibfield  {journal} {\bibinfo  {journal} {IEEE Transactions on Quantum Engineering}\ } (\bibinfo {year} {2023})}\BibitemShut {NoStop}%
\bibitem [{\citenamefont {Fin{\v{z}}gar}\ \emph {et~al.}(2024)\citenamefont {Fin{\v{z}}gar}, \citenamefont {Schuetz}, \citenamefont {Brubaker}, \citenamefont {Nishimori},\ and\ \citenamefont {Katzgraber}}]{finvzgar2024designing}%
  \BibitemOpen
  \bibfield  {author} {\bibinfo {author} {\bibfnamefont {J.~R.}\ \bibnamefont {Fin{\v{z}}gar}}, \bibinfo {author} {\bibfnamefont {M.~J.}\ \bibnamefont {Schuetz}}, \bibinfo {author} {\bibfnamefont {J.~K.}\ \bibnamefont {Brubaker}}, \bibinfo {author} {\bibfnamefont {H.}~\bibnamefont {Nishimori}},\ and\ \bibinfo {author} {\bibfnamefont {H.~G.}\ \bibnamefont {Katzgraber}},\ }\bibfield  {title} {\bibinfo {title} {Designing quantum annealing schedules using {Bayesian} optimization},\ }\href@noop {} {\bibfield  {journal} {\bibinfo  {journal} {Physical Review Research}\ }\textbf {\bibinfo {volume} {6}},\ \bibinfo {pages} {023063} (\bibinfo {year} {2024})}\BibitemShut {NoStop}%
\bibitem [{\citenamefont {Tamiya}\ and\ \citenamefont {Yamasaki}(2022)}]{tamiya2022stochastic}%
  \BibitemOpen
  \bibfield  {author} {\bibinfo {author} {\bibfnamefont {S.}~\bibnamefont {Tamiya}}\ and\ \bibinfo {author} {\bibfnamefont {H.}~\bibnamefont {Yamasaki}},\ }\bibfield  {title} {\bibinfo {title} {Stochastic gradient line {Bayesian} optimization for efficient noise-robust optimization of parameterized quantum circuits},\ }\href@noop {} {\bibfield  {journal} {\bibinfo  {journal} {npj Quantum Information}\ }\textbf {\bibinfo {volume} {8}},\ \bibinfo {pages} {90} (\bibinfo {year} {2022})}\BibitemShut {NoStop}%
\bibitem [{\citenamefont {Kim}\ and\ \citenamefont {Wang}(2023)}]{kim2023quantum}%
  \BibitemOpen
  \bibfield  {author} {\bibinfo {author} {\bibfnamefont {J.~E.}\ \bibnamefont {Kim}}\ and\ \bibinfo {author} {\bibfnamefont {Y.}~\bibnamefont {Wang}},\ }\bibfield  {title} {\bibinfo {title} {Quantum approximate {Bayesian} optimization algorithms with two mixers and uncertainty quantification},\ }\href@noop {} {\bibfield  {journal} {\bibinfo  {journal} {IEEE Transactions on Quantum Engineering}\ } (\bibinfo {year} {2023})}\BibitemShut {NoStop}%
\bibitem [{\citenamefont {Ben{\'\i}tez-Buenache}\ and\ \citenamefont {Portell-Montserrat}(2024)}]{benitez2024bayesian}%
  \BibitemOpen
  \bibfield  {author} {\bibinfo {author} {\bibfnamefont {A.}~\bibnamefont {Ben{\'\i}tez-Buenache}}\ and\ \bibinfo {author} {\bibfnamefont {Q.}~\bibnamefont {Portell-Montserrat}},\ }\bibfield  {title} {\bibinfo {title} {Bayesian parameterized quantum circuit optimization {(BPQCO)}: A task and hardware-dependent approach},\ }\href@noop {} {\bibfield  {journal} {\bibinfo  {journal} {arXiv preprint arXiv:2404.11253}\ } (\bibinfo {year} {2024})}\BibitemShut {NoStop}%
\bibitem [{\citenamefont {Ravi}\ \emph {et~al.}(2022)\citenamefont {Ravi}, \citenamefont {Gokhale}, \citenamefont {Ding}, \citenamefont {Kirby}, \citenamefont {Smith}, \citenamefont {Baker}, \citenamefont {Love}, \citenamefont {Hoffmann}, \citenamefont {Brown},\ and\ \citenamefont {Chong}}]{ravi2022cafqa}%
  \BibitemOpen
  \bibfield  {author} {\bibinfo {author} {\bibfnamefont {G.~S.}\ \bibnamefont {Ravi}}, \bibinfo {author} {\bibfnamefont {P.}~\bibnamefont {Gokhale}}, \bibinfo {author} {\bibfnamefont {Y.}~\bibnamefont {Ding}}, \bibinfo {author} {\bibfnamefont {W.}~\bibnamefont {Kirby}}, \bibinfo {author} {\bibfnamefont {K.}~\bibnamefont {Smith}}, \bibinfo {author} {\bibfnamefont {J.~M.}\ \bibnamefont {Baker}}, \bibinfo {author} {\bibfnamefont {P.~J.}\ \bibnamefont {Love}}, \bibinfo {author} {\bibfnamefont {H.}~\bibnamefont {Hoffmann}}, \bibinfo {author} {\bibfnamefont {K.~R.}\ \bibnamefont {Brown}},\ and\ \bibinfo {author} {\bibfnamefont {F.~T.}\ \bibnamefont {Chong}},\ }\bibfield  {title} {\bibinfo {title} {{CAFQA}: A classical simulation bootstrap for variational quantum algorithms},\ }in\ \href@noop {} {\emph {\bibinfo {booktitle} {Proceedings of the 28th ACM International Conference on Architectural Support for Programming Languages and Operating Systems, Volume 1}}}\ (\bibinfo {year} {2022})\ pp.\ \bibinfo {pages}
  {15--29}\BibitemShut {NoStop}%
\bibitem [{\citenamefont {Cheng}\ \emph {et~al.}(2024)\citenamefont {Cheng}, \citenamefont {Chen}, \citenamefont {Zhang},\ and\ \citenamefont {Zhang}}]{cheng2024quantum}%
  \BibitemOpen
  \bibfield  {author} {\bibinfo {author} {\bibfnamefont {L.}~\bibnamefont {Cheng}}, \bibinfo {author} {\bibfnamefont {Y.-Q.}\ \bibnamefont {Chen}}, \bibinfo {author} {\bibfnamefont {S.-X.}\ \bibnamefont {Zhang}},\ and\ \bibinfo {author} {\bibfnamefont {S.}~\bibnamefont {Zhang}},\ }\bibfield  {title} {\bibinfo {title} {Quantum approximate optimization via learning-based adaptive optimization},\ }\href@noop {} {\bibfield  {journal} {\bibinfo  {journal} {Communications Physics}\ }\textbf {\bibinfo {volume} {7}},\ \bibinfo {pages} {83} (\bibinfo {year} {2024})}\BibitemShut {NoStop}%
\bibitem [{\citenamefont {Sharma}\ \emph {et~al.}(2020)\citenamefont {Sharma}, \citenamefont {Khatri}, \citenamefont {Cerezo},\ and\ \citenamefont {Coles}}]{sharma2020noise}%
  \BibitemOpen
  \bibfield  {author} {\bibinfo {author} {\bibfnamefont {K.}~\bibnamefont {Sharma}}, \bibinfo {author} {\bibfnamefont {S.}~\bibnamefont {Khatri}}, \bibinfo {author} {\bibfnamefont {M.}~\bibnamefont {Cerezo}},\ and\ \bibinfo {author} {\bibfnamefont {P.~J.}\ \bibnamefont {Coles}},\ }\bibfield  {title} {\bibinfo {title} {Noise resilience of variational quantum compiling},\ }\href@noop {} {\bibfield  {journal} {\bibinfo  {journal} {New Journal of Physics}\ }\textbf {\bibinfo {volume} {22}},\ \bibinfo {pages} {043006} (\bibinfo {year} {2020})}\BibitemShut {NoStop}%
\bibitem [{\citenamefont {Nguyen}\ \emph {et~al.}(2020)\citenamefont {Nguyen}, \citenamefont {Gupta}, \citenamefont {Ha}, \citenamefont {Rana},\ and\ \citenamefont {Venkatesh}}]{nguyen2020distributionally}%
  \BibitemOpen
  \bibfield  {author} {\bibinfo {author} {\bibfnamefont {T.}~\bibnamefont {Nguyen}}, \bibinfo {author} {\bibfnamefont {S.}~\bibnamefont {Gupta}}, \bibinfo {author} {\bibfnamefont {H.}~\bibnamefont {Ha}}, \bibinfo {author} {\bibfnamefont {S.}~\bibnamefont {Rana}},\ and\ \bibinfo {author} {\bibfnamefont {S.}~\bibnamefont {Venkatesh}},\ }\bibfield  {title} {\bibinfo {title} {Distributionally robust {Bayesian} quadrature optimization},\ }in\ \href@noop {} {\emph {\bibinfo {booktitle} {International Conference on Artificial Intelligence and Statistics}}}\ (\bibinfo {organization} {PMLR},\ \bibinfo {year} {2020})\ pp.\ \bibinfo {pages} {1921--1931}\BibitemShut {NoStop}%
\bibitem [{\citenamefont {Muandet}\ \emph {et~al.}(2017)\citenamefont {Muandet}, \citenamefont {Fukumizu}, \citenamefont {Sriperumbudur}, \citenamefont {Sch{\"o}lkopf} \emph {et~al.}}]{muandet2017kernel}%
  \BibitemOpen
  \bibfield  {author} {\bibinfo {author} {\bibfnamefont {K.}~\bibnamefont {Muandet}}, \bibinfo {author} {\bibfnamefont {K.}~\bibnamefont {Fukumizu}}, \bibinfo {author} {\bibfnamefont {B.}~\bibnamefont {Sriperumbudur}}, \bibinfo {author} {\bibfnamefont {B.}~\bibnamefont {Sch{\"o}lkopf}}, \emph {et~al.},\ }\bibfield  {title} {\bibinfo {title} {Kernel mean embedding of distributions: A review and beyond},\ }\href@noop {} {\bibfield  {journal} {\bibinfo  {journal} {Foundations and Trends{\textregistered} in Machine Learning}\ }\textbf {\bibinfo {volume} {10}},\ \bibinfo {pages} {1} (\bibinfo {year} {2017})}\BibitemShut {NoStop}%
\bibitem [{\citenamefont {Rasmussen}(2003)}]{rasmussen2003gaussian}%
  \BibitemOpen
  \bibfield  {author} {\bibinfo {author} {\bibfnamefont {C.~E.}\ \bibnamefont {Rasmussen}},\ }\bibfield  {title} {\bibinfo {title} {Gaussian processes in machine learning},\ }in\ \href@noop {} {\emph {\bibinfo {booktitle} {Summer school on machine learning}}}\ (\bibinfo {organization} {Springer},\ \bibinfo {year} {2003})\ pp.\ \bibinfo {pages} {63--71}\BibitemShut {NoStop}%
\bibitem [{\citenamefont {Frazier}(2018)}]{frazier2018tutorial}%
  \BibitemOpen
  \bibfield  {author} {\bibinfo {author} {\bibfnamefont {P.~I.}\ \bibnamefont {Frazier}},\ }\bibfield  {title} {\bibinfo {title} {A tutorial on {Bayesian} optimization},\ }\href@noop {} {\bibfield  {journal} {\bibinfo  {journal} {arXiv preprint arXiv:1807.02811}\ } (\bibinfo {year} {2018})}\BibitemShut {NoStop}%
\bibitem [{\citenamefont {Cakmak}\ \emph {et~al.}(2020)\citenamefont {Cakmak}, \citenamefont {Astudillo~Marban}, \citenamefont {Frazier},\ and\ \citenamefont {Zhou}}]{cakmak2020bayesian}%
  \BibitemOpen
  \bibfield  {author} {\bibinfo {author} {\bibfnamefont {S.}~\bibnamefont {Cakmak}}, \bibinfo {author} {\bibfnamefont {R.}~\bibnamefont {Astudillo~Marban}}, \bibinfo {author} {\bibfnamefont {P.}~\bibnamefont {Frazier}},\ and\ \bibinfo {author} {\bibfnamefont {E.}~\bibnamefont {Zhou}},\ }\bibfield  {title} {\bibinfo {title} {Bayesian optimization of risk measures},\ }\href@noop {} {\bibfield  {journal} {\bibinfo  {journal} {Advances in Neural Information Processing Systems}\ }\textbf {\bibinfo {volume} {33}},\ \bibinfo {pages} {20130} (\bibinfo {year} {2020})}\BibitemShut {NoStop}%
\bibitem [{\citenamefont {Lavrijsen}\ \emph {et~al.}(2020)\citenamefont {Lavrijsen}, \citenamefont {Tudor}, \citenamefont {M{\"u}ller}, \citenamefont {Iancu},\ and\ \citenamefont {De~Jong}}]{lavrijsen2020classical}%
  \BibitemOpen
  \bibfield  {author} {\bibinfo {author} {\bibfnamefont {W.}~\bibnamefont {Lavrijsen}}, \bibinfo {author} {\bibfnamefont {A.}~\bibnamefont {Tudor}}, \bibinfo {author} {\bibfnamefont {J.}~\bibnamefont {M{\"u}ller}}, \bibinfo {author} {\bibfnamefont {C.}~\bibnamefont {Iancu}},\ and\ \bibinfo {author} {\bibfnamefont {W.}~\bibnamefont {De~Jong}},\ }\bibfield  {title} {\bibinfo {title} {Classical optimizers for noisy intermediate-scale quantum devices},\ }in\ \href@noop {} {\emph {\bibinfo {booktitle} {2020 IEEE international conference on quantum computing and engineering (QCE)}}}\ (\bibinfo {organization} {IEEE},\ \bibinfo {year} {2020})\ pp.\ \bibinfo {pages} {267--277}\BibitemShut {NoStop}%
\bibitem [{\citenamefont {Farokhi}(2023)}]{farokhi2023distributionally}%
  \BibitemOpen
  \bibfield  {author} {\bibinfo {author} {\bibfnamefont {F.}~\bibnamefont {Farokhi}},\ }\bibfield  {title} {\bibinfo {title} {Distributionally-robust optimization with noisy data for discrete uncertainties using total variation distance},\ }\href@noop {} {\bibfield  {journal} {\bibinfo  {journal} {IEEE Control Systems Letters}\ } (\bibinfo {year} {2023})}\BibitemShut {NoStop}%
\bibitem [{\citenamefont {Mohajerin~Esfahani}\ and\ \citenamefont {Kuhn}(2018)}]{mohajerin2018data}%
  \BibitemOpen
  \bibfield  {author} {\bibinfo {author} {\bibfnamefont {P.}~\bibnamefont {Mohajerin~Esfahani}}\ and\ \bibinfo {author} {\bibfnamefont {D.}~\bibnamefont {Kuhn}},\ }\bibfield  {title} {\bibinfo {title} {Data-driven distributionally robust optimization using the wasserstein metric: performance guarantees and tractable reformulations},\ }\href@noop {} {\bibfield  {journal} {\bibinfo  {journal} {Mathematical Programming}\ }\textbf {\bibinfo {volume} {171}},\ \bibinfo {pages} {115} (\bibinfo {year} {2018})}\BibitemShut {NoStop}%
\bibitem [{\citenamefont {Srinivas}\ \emph {et~al.}(2010)\citenamefont {Srinivas}, \citenamefont {Krause}, \citenamefont {Kakade},\ and\ \citenamefont {Seeger}}]{srinivas2009gaussian}%
  \BibitemOpen
  \bibfield  {author} {\bibinfo {author} {\bibfnamefont {N.}~\bibnamefont {Srinivas}}, \bibinfo {author} {\bibfnamefont {A.}~\bibnamefont {Krause}}, \bibinfo {author} {\bibfnamefont {S.}~\bibnamefont {Kakade}},\ and\ \bibinfo {author} {\bibfnamefont {M.}~\bibnamefont {Seeger}},\ }\bibfield  {title} {\bibinfo {title} {Gaussian process optimization in the bandit setting: No regret and experimental design},\ }in\ \href@noop {} {\emph {\bibinfo {booktitle} {International Conference on Machine Learning}}}\ (\bibinfo  {publisher} {Omnipress},\ \bibinfo {address} {Madison, WI, USA},\ \bibinfo {year} {2010})\ p.\ \bibinfo {pages} {1015–1022}\BibitemShut {NoStop}%
\bibitem [{\citenamefont {Bogunovic}\ \emph {et~al.}(2018)\citenamefont {Bogunovic}, \citenamefont {Scarlett}, \citenamefont {Jegelka},\ and\ \citenamefont {Cevher}}]{bogunovic2018adversarially}%
  \BibitemOpen
  \bibfield  {author} {\bibinfo {author} {\bibfnamefont {I.}~\bibnamefont {Bogunovic}}, \bibinfo {author} {\bibfnamefont {J.}~\bibnamefont {Scarlett}}, \bibinfo {author} {\bibfnamefont {S.}~\bibnamefont {Jegelka}},\ and\ \bibinfo {author} {\bibfnamefont {V.}~\bibnamefont {Cevher}},\ }\bibfield  {title} {\bibinfo {title} {Adversarially robust optimization with {Gaussian} processes},\ }\href@noop {} {\bibfield  {journal} {\bibinfo  {journal} {Advances in neural information processing systems}\ }\textbf {\bibinfo {volume} {31}} (\bibinfo {year} {2018})}\BibitemShut {NoStop}%
\bibitem [{\citenamefont {Ye}(1998)}]{ye1998orthogonal}%
  \BibitemOpen
  \bibfield  {author} {\bibinfo {author} {\bibfnamefont {K.~Q.}\ \bibnamefont {Ye}},\ }\bibfield  {title} {\bibinfo {title} {Orthogonal column {Latin} hypercubes and their application in computer experiments},\ }\href@noop {} {\bibfield  {journal} {\bibinfo  {journal} {Journal of the American Statistical Association}\ }\textbf {\bibinfo {volume} {93}},\ \bibinfo {pages} {1430} (\bibinfo {year} {1998})}\BibitemShut {NoStop}%
\bibitem [{\citenamefont {Kandala}\ \emph {et~al.}(2017)\citenamefont {Kandala}, \citenamefont {Mezzacapo}, \citenamefont {Temme}, \citenamefont {Takita}, \citenamefont {Brink}, \citenamefont {Chow},\ and\ \citenamefont {Gambetta}}]{kandala2017hardware}%
  \BibitemOpen
  \bibfield  {author} {\bibinfo {author} {\bibfnamefont {A.}~\bibnamefont {Kandala}}, \bibinfo {author} {\bibfnamefont {A.}~\bibnamefont {Mezzacapo}}, \bibinfo {author} {\bibfnamefont {K.}~\bibnamefont {Temme}}, \bibinfo {author} {\bibfnamefont {M.}~\bibnamefont {Takita}}, \bibinfo {author} {\bibfnamefont {M.}~\bibnamefont {Brink}}, \bibinfo {author} {\bibfnamefont {J.~M.}\ \bibnamefont {Chow}},\ and\ \bibinfo {author} {\bibfnamefont {J.~M.}\ \bibnamefont {Gambetta}},\ }\bibfield  {title} {\bibinfo {title} {Hardware-efficient variational quantum eigensolver for small molecules and quantum magnets},\ }\href@noop {} {\bibfield  {journal} {\bibinfo  {journal} {nature}\ }\textbf {\bibinfo {volume} {549}},\ \bibinfo {pages} {242} (\bibinfo {year} {2017})}\BibitemShut {NoStop}%
\bibitem [{\citenamefont {Liu}\ \emph {et~al.}(2022)\citenamefont {Liu}, \citenamefont {Angone}, \citenamefont {Shaydulin}, \citenamefont {Safro}, \citenamefont {Alexeev},\ and\ \citenamefont {Cincio}}]{liu2022layer}%
  \BibitemOpen
  \bibfield  {author} {\bibinfo {author} {\bibfnamefont {X.}~\bibnamefont {Liu}}, \bibinfo {author} {\bibfnamefont {A.}~\bibnamefont {Angone}}, \bibinfo {author} {\bibfnamefont {R.}~\bibnamefont {Shaydulin}}, \bibinfo {author} {\bibfnamefont {I.}~\bibnamefont {Safro}}, \bibinfo {author} {\bibfnamefont {Y.}~\bibnamefont {Alexeev}},\ and\ \bibinfo {author} {\bibfnamefont {L.}~\bibnamefont {Cincio}},\ }\bibfield  {title} {\bibinfo {title} {Layer {VQE}: A variational approach for combinatorial optimization on noisy quantum computers},\ }\href@noop {} {\bibfield  {journal} {\bibinfo  {journal} {IEEE Transactions on Quantum Engineering}\ }\textbf {\bibinfo {volume} {3}},\ \bibinfo {pages} {1} (\bibinfo {year} {2022})}\BibitemShut {NoStop}%
\bibitem [{\citenamefont {Wang}\ \emph {et~al.}(2021{\natexlab{a}})\citenamefont {Wang}, \citenamefont {Czarnik}, \citenamefont {Arrasmith}, \citenamefont {Cerezo}, \citenamefont {Cincio},\ and\ \citenamefont {Coles}}]{wang2021can}%
  \BibitemOpen
  \bibfield  {author} {\bibinfo {author} {\bibfnamefont {S.}~\bibnamefont {Wang}}, \bibinfo {author} {\bibfnamefont {P.}~\bibnamefont {Czarnik}}, \bibinfo {author} {\bibfnamefont {A.}~\bibnamefont {Arrasmith}}, \bibinfo {author} {\bibfnamefont {M.}~\bibnamefont {Cerezo}}, \bibinfo {author} {\bibfnamefont {L.}~\bibnamefont {Cincio}},\ and\ \bibinfo {author} {\bibfnamefont {P.~J.}\ \bibnamefont {Coles}},\ }\bibfield  {title} {\bibinfo {title} {Can error mitigation improve trainability of noisy variational quantum algorithms?},\ }\href@noop {} {\bibfield  {journal} {\bibinfo  {journal} {arXiv preprint arXiv:2109.01051}\ } (\bibinfo {year} {2021}{\natexlab{a}})}\BibitemShut {NoStop}%
\bibitem [{\citenamefont {Wang}\ \emph {et~al.}(2021{\natexlab{b}})\citenamefont {Wang}, \citenamefont {Fontana}, \citenamefont {Cerezo}, \citenamefont {Sharma}, \citenamefont {Sone}, \citenamefont {Cincio},\ and\ \citenamefont {Coles}}]{wang2021noise}%
  \BibitemOpen
  \bibfield  {author} {\bibinfo {author} {\bibfnamefont {S.}~\bibnamefont {Wang}}, \bibinfo {author} {\bibfnamefont {E.}~\bibnamefont {Fontana}}, \bibinfo {author} {\bibfnamefont {M.}~\bibnamefont {Cerezo}}, \bibinfo {author} {\bibfnamefont {K.}~\bibnamefont {Sharma}}, \bibinfo {author} {\bibfnamefont {A.}~\bibnamefont {Sone}}, \bibinfo {author} {\bibfnamefont {L.}~\bibnamefont {Cincio}},\ and\ \bibinfo {author} {\bibfnamefont {P.~J.}\ \bibnamefont {Coles}},\ }\bibfield  {title} {\bibinfo {title} {Noise-induced barren plateaus in variational quantum algorithms},\ }\href@noop {} {\bibfield  {journal} {\bibinfo  {journal} {Nature communications}\ }\textbf {\bibinfo {volume} {12}},\ \bibinfo {pages} {6961} (\bibinfo {year} {2021}{\natexlab{b}})}\BibitemShut {NoStop}%
\bibitem [{\citenamefont {Fontana}\ \emph {et~al.}(2022)\citenamefont {Fontana}, \citenamefont {Cerezo}, \citenamefont {Arrasmith}, \citenamefont {Rungger},\ and\ \citenamefont {Coles}}]{fontana2022non}%
  \BibitemOpen
  \bibfield  {author} {\bibinfo {author} {\bibfnamefont {E.}~\bibnamefont {Fontana}}, \bibinfo {author} {\bibfnamefont {M.}~\bibnamefont {Cerezo}}, \bibinfo {author} {\bibfnamefont {A.}~\bibnamefont {Arrasmith}}, \bibinfo {author} {\bibfnamefont {I.}~\bibnamefont {Rungger}},\ and\ \bibinfo {author} {\bibfnamefont {P.~J.}\ \bibnamefont {Coles}},\ }\bibfield  {title} {\bibinfo {title} {Non-trivial symmetries in quantum landscapes and their resilience to quantum noise},\ }\href@noop {} {\bibfield  {journal} {\bibinfo  {journal} {Quantum}\ }\textbf {\bibinfo {volume} {6}},\ \bibinfo {pages} {804} (\bibinfo {year} {2022})}\BibitemShut {NoStop}%
\bibitem [{\citenamefont {Rabinovich}\ \emph {et~al.}(2022)\citenamefont {Rabinovich}, \citenamefont {Campos}, \citenamefont {Adhikary}, \citenamefont {Pankovets}, \citenamefont {Vinichenko},\ and\ \citenamefont {Biamonte}}]{rabinovich2022gate}%
  \BibitemOpen
  \bibfield  {author} {\bibinfo {author} {\bibfnamefont {D.}~\bibnamefont {Rabinovich}}, \bibinfo {author} {\bibfnamefont {E.}~\bibnamefont {Campos}}, \bibinfo {author} {\bibfnamefont {S.}~\bibnamefont {Adhikary}}, \bibinfo {author} {\bibfnamefont {E.}~\bibnamefont {Pankovets}}, \bibinfo {author} {\bibfnamefont {D.}~\bibnamefont {Vinichenko}},\ and\ \bibinfo {author} {\bibfnamefont {J.}~\bibnamefont {Biamonte}},\ }\bibfield  {title} {\bibinfo {title} {On the gate-error robustness of variational quantum algorithms},\ }\href@noop {} {\bibfield  {journal} {\bibinfo  {journal} {arXiv preprint arXiv:2301.00048}\ } (\bibinfo {year} {2022})}\BibitemShut {NoStop}%
\bibitem [{\citenamefont {Pan}\ \emph {et~al.}(2023)\citenamefont {Pan}, \citenamefont {He}, \citenamefont {Guo},\ and\ \citenamefont {Zhang}}]{he_iccad2023}%
  \BibitemOpen
  \bibfield  {author} {\bibinfo {author} {\bibfnamefont {Y.}~\bibnamefont {Pan}}, \bibinfo {author} {\bibfnamefont {Z.}~\bibnamefont {He}}, \bibinfo {author} {\bibfnamefont {N.}~\bibnamefont {Guo}},\ and\ \bibinfo {author} {\bibfnamefont {Z.}~\bibnamefont {Zhang}},\ }\bibfield  {title} {\bibinfo {title} {Distributionally robust circuit design optimization under variation shifts},\ }in\ \href@noop {} {\emph {\bibinfo {booktitle} {2023 IEEE/ACM International Conference on Computer Aided Design (ICCAD)}}}\ (\bibinfo {organization} {IEEE},\ \bibinfo {year} {2023})\ pp.\ \bibinfo {pages} {1--8}\BibitemShut {NoStop}%
\bibitem [{\citenamefont {Hao}\ \emph {et~al.}(2023)\citenamefont {Hao}, \citenamefont {Liu},\ and\ \citenamefont {Tannu}}]{hao2023enabling}%
  \BibitemOpen
  \bibfield  {author} {\bibinfo {author} {\bibfnamefont {T.}~\bibnamefont {Hao}}, \bibinfo {author} {\bibfnamefont {K.}~\bibnamefont {Liu}},\ and\ \bibinfo {author} {\bibfnamefont {S.}~\bibnamefont {Tannu}},\ }\bibfield  {title} {\bibinfo {title} {Enabling high performance debugging for variational quantum algorithms using compressed sensing},\ }in\ \href@noop {} {\emph {\bibinfo {booktitle} {Proceedings of the 50th Annual International Symposium on Computer Architecture}}}\ (\bibinfo {year} {2023})\ pp.\ \bibinfo {pages} {1--13}\BibitemShut {NoStop}%
\bibitem [{\citenamefont {P{\'e}rez-Salinas}\ \emph {et~al.}(2024)\citenamefont {P{\'e}rez-Salinas}, \citenamefont {Wang},\ and\ \citenamefont {Bonet-Monroig}}]{perez2024analyzing}%
  \BibitemOpen
  \bibfield  {author} {\bibinfo {author} {\bibfnamefont {A.}~\bibnamefont {P{\'e}rez-Salinas}}, \bibinfo {author} {\bibfnamefont {H.}~\bibnamefont {Wang}},\ and\ \bibinfo {author} {\bibfnamefont {X.}~\bibnamefont {Bonet-Monroig}},\ }\bibfield  {title} {\bibinfo {title} {Analyzing variational quantum landscapes with information content},\ }\href@noop {} {\bibfield  {journal} {\bibinfo  {journal} {npj Quantum Information}\ }\textbf {\bibinfo {volume} {10}},\ \bibinfo {pages} {27} (\bibinfo {year} {2024})}\BibitemShut {NoStop}%
\bibitem [{\citenamefont {Hao}\ \emph {et~al.}(2024)\citenamefont {Hao}, \citenamefont {He}, \citenamefont {Shaydulin}, \citenamefont {Pistoia},\ and\ \citenamefont {Tannu}}]{hao2024variational}%
  \BibitemOpen
  \bibfield  {author} {\bibinfo {author} {\bibfnamefont {T.}~\bibnamefont {Hao}}, \bibinfo {author} {\bibfnamefont {Z.}~\bibnamefont {He}}, \bibinfo {author} {\bibfnamefont {R.}~\bibnamefont {Shaydulin}}, \bibinfo {author} {\bibfnamefont {M.}~\bibnamefont {Pistoia}},\ and\ \bibinfo {author} {\bibfnamefont {S.}~\bibnamefont {Tannu}},\ }\bibfield  {title} {\bibinfo {title} {Variational quantum algorithm landscape reconstruction by low-rank tensor completion},\ }\href@noop {} {\bibfield  {journal} {\bibinfo  {journal} {arXiv preprint arXiv:2405.10941}\ } (\bibinfo {year} {2024})}\BibitemShut {NoStop}%
\end{thebibliography}%

\end{document}